%% file: main.tex
\documentclass[twoside,11pt]{article}
\input{header}

\begin{document}
\title{On the Three Demons in Causality in Finance: Time  Resolution, Nonstationarity, and Latent Factors}

   
\author{
Xinshuai Dong$^{1}$ \quad Haoyue Dai$^{1}$ \quad
Yewen Fan$^{1}$ \quad Songyao Jin$^{2}$ \quad
Sathyamoorthy Rajendran$^{2}$ \quad Kun Zhang$^{12}$
\\
$^1$Carnegie Mellon University\\
$^2$Mohamed bin Zayed University of Artificial Intelligence
 }

\editor{My editor}
\maketitle

\input{sec/0_abstract}

\input{sec/1_introduction}

\input{sec/2_aggregation}

\input{sec/3_time_varying}

\input{sec/4_latent_vars}

\input{sec/5_experiments}

\input{sec/7_conclusion_discussion}


\newpage
\appendix

\bibliography{references}

\end{document}

%% file: header.tex
\usepackage[preprint]{jmlr2e}
\usepackage{lastpage}
\ShortHeadings{On the Three Demons: Time Resolution, Nonstationarity,
and Latent Factors}{Dong, Dai, Fan, Jin, Rajendran, and Zhang}
\firstpageno{1}

\usepackage[utf8]{inputenc} %
\usepackage[T1]{fontenc}    %
\usepackage{hyperref}       %
\usepackage{url}            %
\usepackage{booktabs}       %
\usepackage{amsfonts}       %
\usepackage{nicefrac}       %
\usepackage{microtype}      %
\usepackage{xcolor}         %
\usepackage{wrapfig}
\usepackage{tikz}
\usetikzlibrary{arrows,positioning}

\usepackage{nccmath}
\usepackage{amsmath}
\usepackage{amssymb}
\usepackage{bm}
\usepackage{bbm}

\usepackage{stackengine}
\usepackage{subcaption}
\usepackage{paralist}
\usepackage{tabularx}
\usepackage{scalerel,graphicx,xparse,textcomp}
\usepackage{mathtools}

\usepackage{etoolbox}
\usepackage[capitalize,noabbrev]{cleveref}

\usepackage{enumitem}
\usepackage{multirow}
\usepackage{changepage} %
\usepackage{adjustbox} %
\usepackage{makecell}
\usepackage{lipsum}
\usepackage{cancel}
\usepackage{algorithm}
\usepackage{algorithmic}
\usepackage{multicol}
\usepackage{pifont}
\usepackage{stmaryrd}
\usepackage{trimclip}
\usepackage{tikzducks}
\usepackage[normalem]{ulem}
\usepackage{setspace}

\usepackage{float}
\usepackage{verbatim}
\usepackage{array}
\usepackage{xspace}

\let\emptyset\varnothing

\newcommand{\mbX}{\mathbf{X}}
\newcommand{\mbe}{\mathbf{e}}
\newcommand{\tildeX}{\Tilde{\mathbf{X}}}
\newcommand{\tildee}{\Tilde{\mathbf{e}}}

\graphicspath{{img/}}

\newcommand{\node}[1]{\mathsf{#1}}
\newcommand{\set}[1]{\mathbf{#1}}
\newcommand{\setset}[1]{\mathcal{#1}}
\newcommand{\setsetset}[1]{\mathbb{#1}}
\newcommand{\graph}{\mathcal{G}}
\newcommand{\graphp}{\mathcal{G'}}

\newcommand{\parents}{\text{Pa}_{\graph}}

\newcommand{\purechildren}{\text{PCh}_{\graph}}

\newcommand{\meassureddes}{\text{MDe}_{\graph}}

%% file: sec/0_abstract.tex
\begin{abstract}
Financial data is generally time series in essence and thus suffers from three fundamental issues: the mismatch in time resolution, the time-varying property of the distribution - nonstationarity, and causal factors that are important but unknown/unobserved.
In this paper, we follow a causal perspective to systematically look into these three demons in finance. Specifically, we reexamine these issues in the context of causality, which gives rise to a novel and inspiring understanding of how the issues can be addressed. Following this perspective, we provide systematic solutions to these problems, which hopefully would serve as a foundation for future research in the area.
\end{abstract}


%% file: sec/1_introduction.tex
\section{Introduction and Related Work}\label{sec:introduction}




Financial investments through a scientific decision making process is a major concern for both retail and professional investors. Though many literature seem to address this, yet they fail to completely explain the underlying process in a principled way. On the other hand, hedge funds and institutions give less significance to explicitly convey the reasoning behind their investment strategies. While the customers of such funds are satisfied with the profitable performances, it becomes a serious problem when they start to lose money.

To reveal logical reasons, Causality played a compelling role in uncovering highly accurate explanations of several phenomena, especially in health sciences, climate change and energy consumption. Following success in other fields, it is being actively tested in the finance domain recently. The advancements in Causal discovery \citep{spirtes2000causation} ease the procedure of arriving at approximate or better solutions. In the early days, do-calculus introduced by \cite{pearl2009causality} encouraged the researchers to transition the way of thinking with its implicit structure.

In econometrics, the factor models dominated the investment strategies for some time until the investors realized that it did not have a proper scientific foundation. Further, most of the retail traders, as well as institutions mistake that their decisions are based on the cause and effect paradigm while they are actually associational or correlational claims. Many such examples are explained through Type A and Type B Spuriosity in \citep{lopez2022causal}. A strong theory should be able to signal any upcoming black swan events and should provide the scenarios of falsification.

Comprehending the interdependence among time series is pivotal for studying the underlying mechanisms of complex systems. The ability to delineate the causal structure within financial data holds significant potential benefits for financial agencies, particularly stakeholders in the stock market. 
This paper concentrates on three demons in the the causal discovery of financial data: low observed time resolution, non-stationary data, and the existence of latent factors.

Many time series causal discovery methods, such as Granger causality \citep{granger1969investigating}, typically assume that the frequency at which the data is measured aligns with the true causal frequency of the underlying process. However, numerous causal processes unfold rapidly, and the time series data we observe often has a frequency much lower than the causal frequency, usually representing a temporal aggregation of the original causal processes \citep{fisher1970correspondence}. Extensive investigations have indicated that temporal aggregation can introduce errors in the estimation of causal relations \citep{marcellino1999some,breitung2002temporal,rajaguru2008temporal} This raises a fundamental question: how can we leverage low-resolution, temporally aggregated data to unveil meaningful causal relations? 

Another focal point in causal discovery and finance is the issue of data nonstationarity, influencing the outcomes of causal discovery methods applied to financial data.  Nonstationarity implies that causal relations and their strengths vary with time \citep{zhang2015discovery}. Within finance, this phenomenon is commonly referred to as concept drift, signifying the changing distribution of data over time \citep{lu2018learning}. Traditional causal discovery methods such as PC \citep{spirtes2000causation}, GES \citep{chickering2002optimal}, and LiNGAM \citep{shimizu2006linear} were originally designed for the i.i.d. case and did not account for distribution shifts. A recent method, PC-MCI \citep{runge2019detecting}, is specifically tailored for time series data; however, it still exhibits limitations when confronted with changing causal relations over time.

Moreover, in financial data analysis, latent factors refer to unobservable variables that underlie the observed market dynamics and influence multiple observable variables simultaneously \citep{lopez2022causal}. The concept of latent factors is pivotal in causal discovery within the financial domain as it captures the latent structures and commonalities inherent in complex financial systems. The potential bad effect arises when the true causal relationships between observed variables are obscured by the influence of latent factors, introducing spurious correlations and complicating the accurate identification of causal structures \citep{spirtes2000causation}. Consequently, latent factors demand careful consideration and rigorous methods in causal discovery processes to distinguish true causal relationships from those induced by the latent variables, ensuring the robustness and reliability of the discovered results in financial data analysis. 

The goal of this paper is to bring attentions into the three demons in the intersection of causality and finance data analysis, presenting systematic solutions from a causal perspective. In Section 1.1, we provide the requisite background concepts in causal discovery. Section 2 delves into the issue of low resolution and temporal aggregated data, offering a meticulous mathematical analysis that demonstrates, under specific technical assumptions, the potential unveiling of causal relations through the direct application of temporal aggregated data to instantaneous causal discovery methods. In Section 3, we explain the nonstationary property of data distribution from both financial and causal perspectives, introducing our proposed method, CD-NOD, which strategically considers time as a confounding factor. Section 4 scrutinizes the adverse effects of latent causal factors, subsequently introducing our recent work involving rank deficiency methods. Our empirical findings on causal relations among select S$\&$P 100 stocks are presented in Section 5. Finally, Section 6 provides a succinct discussion and conclusion, consolidating our insights into a comprehensive framework for navigating the intricate landscape of causality and financial data analysis.

Our contributions can be summarised as below: 
\begin{itemize}
  \item We delineate three pivotal issues concerning causality in financial data, i.e.,  time resolution, nonstationarity, and latent variables, and undertake a comprehensive exploration on these challenges following a systematic and causal point of view.
  \item We provide a novel and inspiring causal understanding of the three issues. Based on this understanding we propose rigorous methods to address each issue as well as a   systematic solution that works even when the three issues are present simultaneously.
  
  \item Our empirical study validates the effectiveness of our proposed methods in real-life scenarios in finite sample cases. Our techniques and findings hopefully shed new light on the problem and serve as a foundation for further exploration in the area.
\end{itemize}


\subsection{Concepts and Terminology in Causal Discovery}


To furnish a thorough introduction to causal discovery methods applied to financial data in the subsequent sections, we present key concepts within the field of causal discovery below.

\textbf{Structural Causal Model.} The Structural Causal Model (SCM) \citep{pearl2009causality} comprises a set of endogenous variables ($V$) and a set of exogenous variables ($U$), interconnected by a set of functions ($F$) that illustrate how the changes of exogenous variables $U$ impact the values of endogenous variables $V$ in the data generation process. For instance, for $x_i \in V$, $u_i \in U$, and $f_i \in F$, the model $x_i := f_i(Pa(x_i), u_i), i = 1,\dots,d$, signifies the assignment of the value $x_i$ to a function $f_i$ of its structural parents $Pa(x_i)$ and exogenous variable $u_i$. For each SCM, a causal graph $\mathcal{G}$ is correspondingly generated by representing each $x_i$ with each vertex and adding directing edges from each parent variable in $Pa(x_i)$ to the child $x_i$.


\textbf{Causal Sufficiency.} A set of observed variables is deemed causally sufficient if all common causes of these variables are also observed \citep{spirtes2000causation}. In other words, there is no latent confounder in the graph and all causally related variables are observed.

\textbf{Causal Markov Condition.} In the causal graph, each variable is independent of its non-descendants conditional on all of its direct causes \citep{hausman1999independence}. This relationship establishes a connection from the causal graph to the statistical properties.

\textbf{Markov Equivalence Class.} A set of directed acyclic graphs belong to the same Markov Equivalent Class if they entail the same conditional independence relations among the observed variables \citep{pearl2000models,spirtes2000causation}. Elements in the same class have the same causal skeleton and v-structures.

\textbf{Faithfulness.} A probability distribution P is faithful to a causal graph $\mathcal{G}$ if all conditional independence relations in the probability distribution P are entailed by the causal graph $\mathcal{G}$ \citep{spirtes2000causation}. This relationship establishes a connection from the statistical properties to the causal graph.

%% file: sec/2_aggregation.tex
\section{Time Resolution and Aggregation}
\label{sec:time_resolution}

It is widely acknowledged that causation unfolds over time, with reactions manifesting not instantly but with a slight time lag, prompting the analysis of time series data. Time series data are typically collected at fixed frequencies. However, numerous causal processes transpire too rapidly for direct observation \citep{fisher1970correspondence}. The data presented to us lack the temporal granularity to capture such swift reactions; instead, what we observe is usually a temporal aggregation of the original causal processes. This situation is prevalent in the financial field. For instance, in the stock market, where changes occur at second intervals, log returns are aggregated at daily, hourly, and minute scales \citep{narang2013inside}. 

The detrimental effects of temporal aggregation on time series models have been examined in previous studies \citep{tiao1972asymptotic,weiss1984systematic}, and subsequent investigations have revealed that temporal aggregation can introduce errors in the estimation of causal relations \citep{marcellino1999some,breitung2002temporal,rajaguru2008temporal}. Classical time series causal discovery methods, such as Granger causal analysis \citep{granger1969investigating}, exhibit sensitivity to temporal aggregation\citep{breitung2002temporal}. 

Numerous econometricians and statisticians have endeavored to recover high-frequency time series by employing disaggregation techniques on low-frequency time series. \citet{stram1986methodological} conducted disaggregation by reinterpreting the generalized least squares problem as a time series task applied to a series of time series totals. \citet{harvey2000estimating} established an appropriate time series model in state-space form and computed the underlying instantaneous change using the Kalman filter. \citet{proietti2006temporal} advocated the application of state space methods, particularly dynamic regression models, as a means to address the challenges associated with temporal disaggregation.

While substantial efforts have been devoted to the disaggregation of time series data, the endeavor to estimate causal structures from temporally aggregated time series remains relatively limited. Fortunately, recent demonstrations indicate that, in the context of a linear causal process,when the number of raw data points forming each observed, averagely aggregated data point is sufficiently large, the observed time series exhibits behaviors akin to independence and identical distribution (i.i.d.) (instantaneous) status \citep{gong2017causal}. This allows for the direct application of instantaneous causal discovery methods to shed light on underlying true time-delayed causal relations.

Let us consider such a causal process modeled by typical VAR model:

\begin{equation}
\label{eq1}
\mbX_t = \mathrm{A} \mbX_{t-1} + \mbe_t,
\end{equation}

where $\mbX_t = (X_{t,1}, \dots, X_{t,n})^T$ represents the original data vector, $\mbe_t = (e_{t,1},  \dots,e_{t,n})^T$ is the temporally and contemporaneously independent noise vector, and $\mathrm{A}$ is the causal transition matrix. 
Furthermore, the temporally aggregated data (observed data)
\begin{equation}
\label{eq2}
\begin{aligned}
\tildeX_t = \frac{1}{k}\sum_{i=1}^k \mbX_{i+(t-1)k}
\end{aligned}
\end{equation}
is obtained by taking the average of every successive k data points. Thus, we have temporally aggregated time series $\tildeX_{1:T} = (\tildeX_{1}, \tildeX_{2}, \dots, \tildeX_{T})$.

Taking Eq. \ref{eq1} and Eq. \ref{eq2} together, we have:

\begin{equation}
\begin{aligned}
\tildeX_t &= \frac{1}{k}\sum_{i=1}^k \mbX_{i+(t-1)k}
          = \frac{1}{k}\sum_{i=1}^k \left(\mathrm{A} \mbX_{i+(t-1)k-1} + \mbe_{i+(t-1)k}\right) \\
          &= \mathrm{A}\left[ \frac{1}{k} \sum_{i=1}^k \mbX_{i+(t-1)k} - \frac{1}{k} \left( \mbX_{tk} - \mbX_{(t-1)k} \right) \right] + \frac{1}{k}\sum_{i=1}^k \mbe_{i+(t-1)k}.
\end{aligned}
\end{equation}

For the high-frequency causal process, it is definitely that the aggregation factor $k$ is very large. As $k \rightarrow \infty$, $\frac{1}{k} \left( \mbX_{tk} - \mbX_{(t-1)k} \right) \rightarrow 0$. Therefore, as $k \rightarrow \infty$, we have
\begin{equation}
\label{eq4}
\begin{aligned}
\tildeX_t = \mathrm{A} \tildeX_t + \tildee_t,
\end{aligned}
\end{equation}
where $\tildee_t = \frac{1}{k}\sum_{i=1}^k \mbe_{i+(t-1)k}$ is the mixture of independent component vectors $\mbe_t$, and elements in $\tildee_t$ are contemporaneous independent to each other. Clearly, $\tildeX_t$ exhibits behaviors resembling a linear instantaneous causal process in the i.i.d. case, and hence instantaneous causal discovery methods can be directly applied to aggregated data to discover meaningful causal relationships (detailed in Section~\ref{sec:latent}).



%% file: sec/3_time_varying.tex
\section{Nonstationary Relationships}

\subsection{From the Financial Perspective}
It is well-known that financial markets has a non-stationary nature. For instance, substantial evidence suggests regime shifts in the crude oil market, as highlighted by \citep{vo2009regime}. Financial market time series data are prone to undergo unpredictable changes, which is known as concept drift \citep{lu2018learning, vzliobaite2016overview, gama2014survey}.

Concept drift is a phenomenon where the distribution of streaming data undergoes possibly unpredictable changes over time, as described in \citep{lu2018learning}. In formal terms, when the feature vector is represented by $x$ and the label by $y$, concept drift is characterized by the time-dependent alteration in the joint distribution of $p(x, y)$, as detailed in \citep{lu2014concept}.

\begin{figure}[t]
\center
    \includegraphics[width = \textwidth]{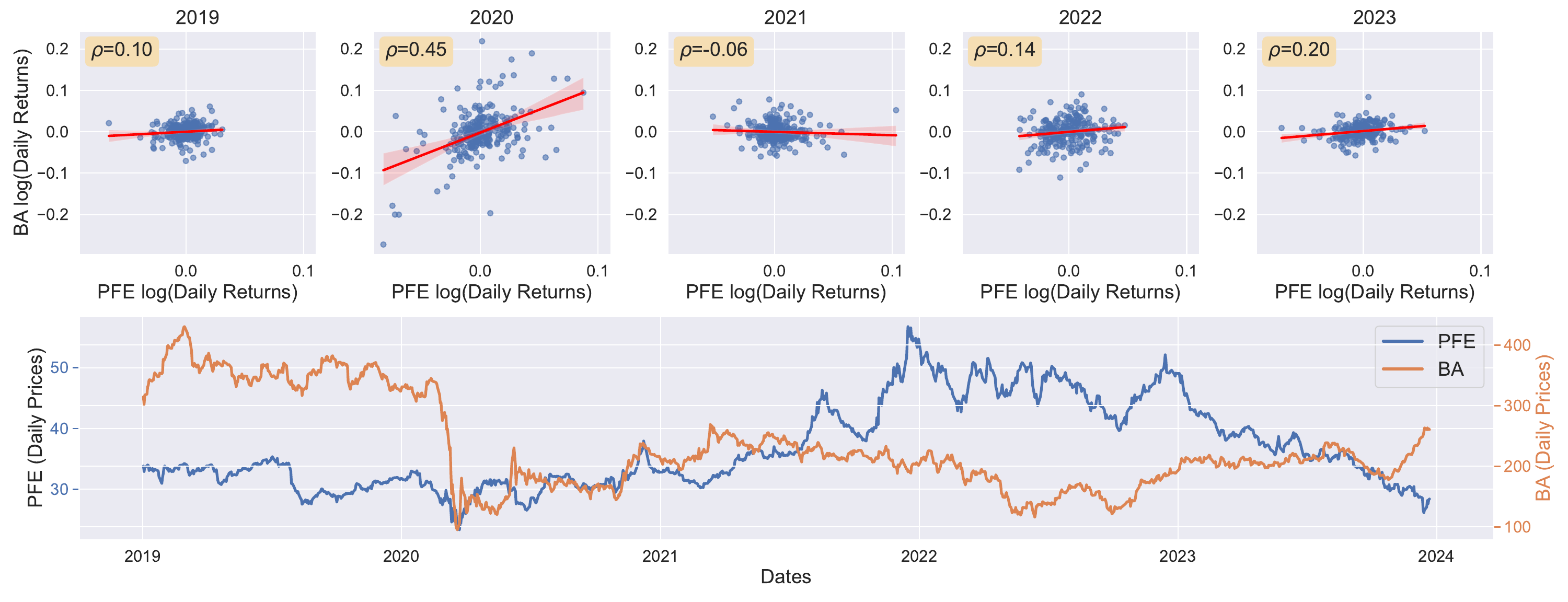}
    \caption{Example to illustrate the nonstationary relationship between Pfizer Inc. (PFE) and The Boeing Company (BA). The upper row: five scatter plots correspond to years from 2019 to 2023. The scatters are the log daily returns of the two stocks. Clearly they have different dependence patterns. The linear regression line and the correlation are annotated. The lower row: daily prices series of the two stocks are plotted for reference.}

    \label{fig:scatter_plot}
\end{figure}

\begin{example}
\label{example1}

Figures \ref{fig:scatter_plot} demonstrates the evolving correlation between the stock price daily log return time series of Pfizer Inc. (PFE) and The Boeing Company (BA) from 2019 to 2023. This relationship has experienced significant shifts, particularly during the pre- and post-COVID periods. In 2019, a time of economic expansion, both PFE and BA exhibited an upward trend in stock prices. The correlation observed in this era could be influenced by broader economic factors (confounders). The correlation is relatively weak in this period. The onset of COVID-19 in early 2020 saw a sharp decline in both BA's and PFE's stock prices. However, in late 2020, both stocks generally increased, displaying a notably stronger correlation (also demonstrated by the correlation between the log daily returns in 2020). This change is possibly linked to Pfizer's vaccine development positively impacting travel confidence, thereby affecting the aviation sector. After 2021, this correlation moderated, aligning with the evolving COVID-19 situation. By 2023, as COVID-19 concerns eased, the correlation between PFE and BA stock prices significantly weakened.

Of course, it's worth noting that pairwise relationships don't always reflect true causality (may be subject to confounders), which is typically captured by conditional relationships. Therefore, the explanations provided above serve more for intuitive understanding than accurate discovery. For a more systematic investigation of this example, please refer to~\cref{sec:experiments}.

\end{example}

To handle concept drift, prior works usually rely on the assumption that the latest data contains more useful information. Researchers have proposed to use retraining based methods \citep{bach2008paired}, ensemble based methods \citep{gomes2017adaptive}, incremental adaptation based methods \citep{domingos2000mining} to tackle concept drift for the general time series tasks. In recent developments, new approaches have been introduced specifically for financial applications. \citet{li2022ddg} suggests initially training a model to predict future data distributions. This model is then used to create training examples that mirror the anticipated future data, which are subsequently utilized to train machine learning models. On the other hand, \citet{riva2022addressing} advocates for treating the trading task as a non-stationary reinforcement learning problem and selects the most effective strategy from a set of reinforcement learning models.

\subsection{From Causal Discovery Perspective}

The profound link between financial data analysis and causal discovery was introduced in~\cref{sec:introduction}. This raises an intriguing question: how should we approach nonstationarity in the context of causal discovery? Is this a obstacle to overcome, or can it offer unexpected advantages?

On one hand, conventional causal discovery methods designed for i.i.d. samples (e.g., PC~\cite{spirtes2000causation} and GES~\citep{chickering2002optimal}), or other fixed functional causal-based methods (e.g., LiNGAM), will give incorrect results with nonstationarity. These methods falter because the conditional independence relationships and the functional constraints in data with shifting distributions differ from those implied by the true causal structure. Even methods tailored for time-series data (e.g., PC-MCI~\citep{runge2019detecting}), also fail to identify the correct causal structures due to nonstationarity affecting not just individual time series, but also the evolving causal relationships between them. A potential solution involves Bayesian change point detection to segment the data into approximately stationary time frames for causal discovery on each frame. However, this approach is limited with the assumption of non-continuous change and the focus of changing point detection is on marginal or joint distributions rather than conditional ones with causal interpretations.

On the other hand, it can be noted that concept drift in financial data (i.e., distributional change) and causal models are heavily coupled: causal models offer a compact way to describe data-generating processes \citep{huang2020causal}. From a causal perspective, the concept drift often stems from alterations in a small subset of variables within their causal generating processes over time, as the joint distribution can be factorized into conditional distributions of variables' causal generating processes, and the joint drift can be attributed to changes in the causal mechanisms of a few variables (the so-called minimal-change principle, as another way to put faithfulness). This understanding sheds light on the mechanics behind concept drift and is pivotal in pinpointing the primary factors driving these shifts. Our method, Constraint-based Causal Discovery from Heterogeneous/Nonstationary Data (CD-NOD), elaborated in \citet{huang2020causal}, leverages this insight to unravel these complex dynamics. More interestingly, nonstationary relationships are not necessarily obstacles to be overcome. On the contrary, they can actually provide more information to identify causal directions which might remain unidentifiable with i.i.d. data.

CD-NOD addresses the following three questions: 1. ``How can we efficiently identify variables with changing local mechanisms and reliably recover the skeleton of the causal structure over observed variables?'', 2. ``How can we take advantage of the information carried by distribution shifts for the purpose of identifying causal directions?'', and 3. ``How can we extract from data a low-dimensional and potentially interpretable representation of changes, the so-called ``driving force'' of changing causal mechanisms?''~\citep{huang2020causal}. Regarding question 1, CD-NOD introduces a surrogate variable (which, in the case of finance data here, is the time index) to characterize hidden unknown quantities that lead to the changes over time. By involving this surrogate variable into constraint-based causal discovery methods together with all other variables, the true causal skeleton can be correctly identified, because this surrogate variable can explain all the nonstationarity, and thus by conditioning on it, the (conditional) independence relationships between observed variables are the same as those implied by the true causal structure. Regarding Question 2, the modularity property (also known as the independent change property) of causal systems~\citep{pearl:88} is exploited, i.e., when there are no confounders for $\texttt{cause}$ and $\texttt{effect}$, $P(\texttt{cause})$ and $P(\texttt{effect} | \texttt{cause})$ are either fixed or change independently. Regarding Question 3, note that the variables adjacent to the time index surrogate are exactly the ones that have changing causal mechanisms over time. With the causal structure and the changing causal modules available, CD-NOD finds a low-dimensional and interpretable mapping of the density of a variable given its parental variables and the surrogate index in a nonparametric way, which exactly captures the nonstationarity of the causal mechanism of this variable.

Here we provide only a general overview of CD-NOD to exploit nonstationarity for causal discovery. For more technical details, please refer to~\citep{huang2020causal}. 

%% file: sec/4_latent_vars.tex
\section{Finding Causal Model in the Presence of Latent Factors}
\label{sec:latent}

\subsection{Factor Model from A Causal Perspective}

In stock market prediction, it has been shown by Fama and French \cite{fama1996multifactor} that, their classical 3-factor model well summarizes the cross section of average stock returns as of the mid-1990s.
Over the past 2 decades, researchers have been devoted to refining the 3-factor model, with prominent ones like the 4-factor model \cite{hou2015digesting} and the 5-factor model \cite{fama2015five}.
Though this line of research works well in some scenarios \citep{hou2017comparison},
 it has become clearer recently that it fails to account for a wide array
of asset pricing anomalies \citep{hou2015digesting}.

One can continuously add more factors to enhance the model, and yet, we argue that, without a fundamental causal understanding of the problem, one can never make reliable and interpretable predictions based on a factor model. 
The reason lies in that the real objective of a factor model, say, $\node{Y}=a\node{X}+b\node{Z}+\epsilon$,
is to predict the causal effect $\mathbb{E}[\node{Y}|\text{do}(\node{X})]$, rather than the conditional expectation $\mathbb{E}[\node{Y}|\node{X},\node{Z})]$ \citep{lopez2022causal}.
Therefore, without a causal perspective,
one cannot even correctly formulate the quantity that we really care about.
Even though we might correctly formulate the problem,
it is still very hard, if not impossible, to predict the causal effect, as generally we only have access to observational data.

Fortunately, with advanced causal discovery methods at hand, one can identify the causal graph and then perform 
simulated interventions (e.g., backdoor adjustment\citep{pearl2009causality}) to estimate the causal effect, by making use of only observational data. 
Typical causal discovery methods include 
the constraint-based PC algorithm~\citep{spirtes2000causation} and score-based GES~\citep{chickering2002optimal},
both of which can asymptotically arrive at the Markov equivalence class of the underlying causal graph, by assuming causal sufficiency.

However, latent variables are ubiquitous in finance data, and thus causal sufficiency generally fails to hold in the area.
Roughly speaking, latent variables/factors are those variables that are essential in the underlying causal dynamics, but cannot be directly observed, or even cannot be semantically well captured by existing concepts in the field.
For example, policies taken by governments can have a significant impact on a financial system, and yet some policies may not always be directly observable.
Therefore, causal discovery methods that work in the presence of latent variables would be very favorable.
One famous line of thought that allows latent variables is FCI \citep{spirtes2000causation}
 and its variants~\citep{spirtes2013causal, colombo2012learning, akbari2021recursive}.
It makes full use of CI information and correctly identifies causal relations between observed variables.

Although FCI  provides correct graphical information in the presence of latent variables, its results tend to be too generalized due to large indeterminacies. E.g., whenever it is possible for two measured variables to be confounded, it will indicate so. Moreover, it focuses solely on the causal relations among observed variables, without considering the relations among latent variables. 
Recent advances in overcomplete ICA~\citep{hoyer2008estimation, salehkaleybar2020learning} also aim at solving this problem, but they  have to assume all latent variables are independent and they do not provide unique solutions in general.

Therefore, a  question naturally arises. If we allow all the variables including observed and latent ones to be very flexibly related (e.g., a latent variable can serve as the effect, the cause, or the mediation of any variables), can we still identify the whole underlying causal structure (including the location, cardinality, and relations among all the variables), and thus gain a better understanding of the underlying causal model?
The answer is yes and we shall detail it in the following section.

\subsection{Causal Discovery in the Presence of Latent Variables}

Latent variables are ubiquitous and essential in many fields. For example, pixels of an image are dependent and yet it can be argued that they are confounded by the latent underlying concept. Other examples would be causally related concepts in psychological studies: we are interested in the human hidden mental factors and yet we can only rely on questionnaires to measure the latent mental states.
Therefore, there is an urgent need for a computational method that can learn true causality including causally related latent variables and causal relations. 

Although it is well known to be an extremely hard problem,
interestingly, our recent preliminary results have shown it can be reliably achieved by exploring certain distributional properties of measured variables - by properly making use of rank deficiency and generalized independence noise (GIN) constraints.
Specifically
 one can recover latent variables and involve the causal relations by making use of rank deficiency of covariance matrix \citep{dong2023versatile} or testing for independence between combinations of measured variables known as GIN \citep{xie2020generalized}. 
We provide our solution (more details can be found in our manuscript \citep{dong2023versatile}), and start with the definition of latent linear causal model as follows.

\begin{definition} (Latent Linear Causal Models)
  Suppose a directed acyclic graph
  where each variable $V_i$ is generated following a causal structural model with n+m variables as follows.
\begin{align}
\node{V}_i=\sum \nolimits_{\node{V}_j \in \parents(\node{V}_i)} a_{ij} \node{V}_j + \varepsilon_{\node{V}_i},
\end{align}
where out of n+m variables, n are observed and m are hidden, i.e., $\{{V}_i\}_{i=1}^{m+n}=\{{X}_i\}_{i=1}^{n}\cup\{{L}_i\}_{i=1}^{m}$, and 
$\varepsilon_{{V}_i}$ is the independent noise term.
\label{definition:lcm}
\end{definition}

\textbf{Goal.} In the above definition, we have $n+m$ variables in the causal model and only n out of n+m are observed variables.
Given i.i.d. samples of only observed variables $\set{X}_{\mathcal{G}} := \{\node{X_i}\}_{i=1}^n$,  our goal is to locate all the latent variables $\set{L}_{\mathcal{G}}=\{{L}_i\}_{i=1}^{m}$, determine their cardinalities, and identify the causal structure $\mathcal{G}$ over all the variables, including both latent and observed ones.
To this end, we will leverage the information from the rank of the covariance matrix of observed variables and will assume rank faithfulness for latent linear causal models as follows. A probability distribution $p$ is rank faithful to $\mathcal{G}$ if every rank constraint on a sub-covariance matrix that holds in $p$ is entailed by every linear structural model with respect to $\mathcal{G}$. 
It is the classical faithfulness assumption that is critical and prevalent in causal discovery \citep{spirtes2000causation,huang2022latent,dong2023versatile};
  it holds generically on infinite data, as the set of values of the SCM's free parameters 
  for which rank is not faithful is of Lebesgue measure 0   \citep{spirtes2013calculation-t-separation}.

\textbf{Rank Constraint and Motivation.} We propose to leverage information from the rank of the cross-covariance matrix over observed variables to unveil the underlying causal structure. The intuition lies in that t-separations are good graphical indicators and they can be inferred from rank constraints. 
We begin with a motivating example as follows.

\begin{example}
Suppose we observe four variables $\node{X_1}, \node{X_2}, \node{X_3}, \node{X_4}$ as common children of a latent variable $\node{L_1}$. They adhere to a LLCM (Definition~\ref{definition:lcm}) as follows:

\makeatletter
\let\par\@@par
\par\parshape0
\everypar{}
\begin{wrapfigure}[0]{r}{0.4\textwidth}
\hspace{3em}\begin{tikzpicture}[inner sep=2pt, >=stealth]
\node (X1) {$\node{X_1}$};
\node (X2) [right=10pt of X1] {$\node{X_2}$};
\node (X3) [right=10pt of X2] {$\node{X_3}$};
\node (X4) [right=10pt of X3] {$\node{X_4}$};
\node (L) [above=25pt of $(X2)!0.5!(X3)$] {$\node{L_1}$};
\path [->] (L) edge node[above=2pt]{\scriptsize{$a$}} (X1);
\path [->] (L) edge node[above=4pt,pos=0.9]{\scriptsize{$b$}} (X2);
\path [->] (L) edge node[above=4pt,pos=0.9]{\scriptsize{$c$}} (X3);
\path [->] (L) edge node[above=2pt]{\scriptsize{$d$}} (X4);
\end{tikzpicture}
\end{wrapfigure}

\begin{equation*}
\begin{aligned}
\node{X_1} &= a \node{L_1} + \epsilon_1 \\
\node{X_2} &= b \node{L_1} + \epsilon_2 \\
\node{X_3} &= c \node{L_1} + \epsilon_3 \\
\node{X_4} &= d \node{L_1} + \epsilon_4
\end{aligned}
\end{equation*}
The classical line of thought for causal discovery is to use CI tests to find d-separations in the graph. E.g., if we assume that $\node{L_1}$ is observed,
we can simply get $\node{X_1}\perp \!\!\! \perp\node{X_2}|\node{L_1}$ and thus conclude the skeleton among these three variables. However, 
$\node{L_1}$ is hidden, and thus we cannot find any CI relations from observational data. by using CI information we can only arrive at a fully connected graph among 
$\node{X_1}, \node{X_2}, \node{X_3}, \node{X_4}$.
\label{example2}
\end{example}

By the above example, we show that in the presence of latent variables, CI relations only provide limited information about the underlying graph.
Next, we will show that by using t-separation implied by rank constraint, we are able to identify the latent variable, which is beyond what CI can do. We shall begin with the definition of trek and t-separation.

\begin{definition} [Treks \citep{sullivant2010trek}]
   A trek from $\node{X}$ to $\node{Y}$ is an ordered pair of directed paths 
   $(P_1,P_2)$ where $P_1$ has a sink $\mathsf{X}$, 
   $P_2$ has a sink $\mathsf{Y}$,
    and both $P_1$ and $P_2$ have the common source $\mathsf{Z}$.
\end{definition}

\begin{definition} [T-separation \citep{sullivant2010trek}]
Let $\mathbf{A}$, $\mathbf{B}$, $\mathbf{C}_{\mathbf{A}}$, 
and $\mathbf{C}_{\mathbf{B}}$ be four subsets of $\mathbf{V}_{\mathcal{G}}$ 
in graph $\mathcal{G}$ (not necessarily disjoint). 
($\mathbf{C}_{\mathbf{A}}$,$\mathbf{C}_{\mathbf{B}}$) t-separates $\mathbf{A}$ from $\mathbf{B}$ if for every trek ($P_1$,$P_2$) from a vertex in $\mathbf{A}$ to a vertex in $\mathbf{B}$, either $P_1$ contains a vertex in  $\mathbf{C}_{\mathbf{A}}$ or $P_2$ contains a vertex in  $\mathbf{C}_{\mathbf{B}}$. 
\label{definition:t-sep}
\end{definition}

Readers who are less familiar with t-separation may refer to our manuscript \citep{dong2023versatile} for illustrative examples. Just as d-separation can be implied by CI,
t-separation can be implied by rank of the cross-covariance matrix over observed variables, as follows.

\begin{theorem} [Rank and T-separation \citep{sullivant2010trek}]
    Given two sets of variables $\set{A}$ and $\set{B}$ from a linear model with  $\graph$, 
  $\text{rank}(\Sigma_{\set{A},{\set{B}}}) = 
  \min \{|\set{C}_{\set{A}}|+
  |\set{C}_{\set{B}}|:(\set{C}_{\set{A}},\set{C}_{\set{B}})~
  \\
  \text{t-separates}~\set{A}~\text{from}~\set{B}~\text{in}~\graph\}$, where $\Sigma_{\set{A},{\set{B}}}$ is the cross-covariance over $\set{A}$ and $\set{B}$.
\label{theorem:rank and t}
\end{theorem}

Now we show why t-separation implied by rank is a good graphical indicator, especially in the presence of latent variables.
Again consider the graph in Example~\ref{example2}.
If we can
observe $\node{L_1}$ then 
we can directly do the CI test conditional on $\node{L_1}$, and thus find  the skeleton.
However, checking that CI relation is impossible, and by CI tests we can only find a clique among $\node{X_1},\node{X_2},\node{X_3},\node{X_4}$.
Fortunately, t-separation implied by rank constraints can achieve a similar goal when $\node{L}$ is not observed. 
Specifically, we can observe that 
$\text{rank}(\Sigma_{\{\node{X_1},\node{X_2}\},
\{\node{X_3},\node{X_4}\}})=1$,
which implies the existence of one variable being in the treks between $\{\node{X_1},\node{X_2}\}$ and 
$\{\node{X_3},\node{X_4}\}$.
The rationale behind is that the t-separation of $\set{A}$, $\set{B}$ by $(\set{C}_\set{A},
\set{C}_\set{B})$ can be deduced through rank without observing any element in $(\set{C}_\set{A},\set{C}_\set{B})$.

On the other hand, if we cannot observe $\node{X_4}$, we cannot identify the existence of latent variable $\node{L_1}$.
The reason lies in that if we cannot observe $\node{X_4}$, we cannot calculate $\text{rank}(\Sigma_{\{\node{X_1},\node{X_2}\},
\{\node{X_3},\node{X_4}\}})$. Instead, we can only test $\text{rank}(\Sigma_{\{\node{X_1},\node{X_2}\},
\{\node{X_3}\}})$ and it will always be 1, no matter whether there is a latent variable or not. Therefore, we have the rough intuition that, for a latent variable to be identifiable, it must have enough observed children and neighbors. 
Next, we formalize this intuition into the concept of atomic cover (a cover is a set of variables) as the minimal identifiable substructure in a graph, with effective cardinality of a set of covers $\setset{V}$ defined
as $||\setset{V}||=|(\cup_{\set{V} \in \setset{V}} \set{V})|$, and that $\set{Y}$ are pure children of $\set{X}$, i.e., $\set{Y}\in\purechildren(\set{X})$, iff $\parents(\set{Y}) = \cup_{\node{Y_i} \in \set{Y}} \parents(\node{Y_i}) = \set{X}$  and $\set{X} \cap \set{Y}=\emptyset$.

\begin{definition} [Atomic Cover]
  Let $\set{V}$ be a set of variables in $\graph$ with $|\set{V}|=k$, where $t$ of the $k$ variables are observed, and the remaining $k-t$ are latent. 
  $\set{V}$ is an atomic cover if $\set{V}$ contains a single observed variable (i.e.,  $k=t=1$ ), or 
    if the following conditions hold:
  \begin{itemize}[]
      \item [(i)] There exists 
      a set of atomic covers $\setset{C}$, with $||\setset{C}||\geq k+1-t$, such that
      $\cup_{\set{C} \in \setset{C}} \set{C}\subseteq \purechildren(\set{V})$
      and 
      $\forall \set{C_1}, \set{C_2} \in \setset{C}, \set{C_1}\cap\set{C_2}=\emptyset$

     \item [(ii)] There exists a  set  of covers $\setset{N}$, with $||\setset{N}||\geq k+1-t$,
     such that every element in $\cup_{\set{N} \in \setset{N}} \set{N}$ is a neighbour of $\set{V}$  and 
     $ (\cup_{\set{N} \in \setset{N}} \set{N}) \cap (\cup_{\set{C} \in \setset{C}} \set{C})=\emptyset$.

\item [(iii)] There is not a partition of $\set{V}= \set{V_1} \cup \set{V_2}$, s.t., both $\set{V_1}$ and  $\set{V_2}$ 
are atomic covers.
  \end{itemize}
  \vspace{-0.5mm}
\label{definition:ac}
\end{definition}

The concept of atomic cover is essential as it is the minimal identifiable substructure in a graph. 
Specifically, given the following Condition~\ref{cond:basic}, we have that the rank deficiency property uniquely implies an atomic cover, formally captured by Theorem~\ref{thm:URD}.

\begin{condition}[Basic Graphical Conditions for Identifiability] A graph $\graph$ satisfies the basic graphical condition for identifiability, if 
 $\forall \node{L}\in\set{L}_\graph$,  $\node{L}$ belongs to at least one atomic cover in $\graph$ and no variable is involved in any triangle structure.
\label{cond:basic}
\end{condition}

\begin{theorem}
 [Uniqueness of Rank Deficiency]
   Suppose $\mathcal{G}$ satisfies Condition~\ref{cond:basic}.
  We further assume  
  (i) all the atomic covers with cardinality $k'<k$ have been discovered and recorded, and (ii) there is no collider
  in $\graph$. 
  If there exists a
  set of observed variables $\set{X}$ and a set of atomic covers $\setset{C}$ satisfying
$\setset{X}=\text{Sep}(\set{X})=\cup_{\node{X} \in \set{X}} \{\{\node{X}\}\}$, $\setset{C}\cap\setset{X}=\emptyset$,  and 
   $||\setset{C}||+||\setset{X}||=k+1$,
   such that (i) For all recorded $k'$ cluster $\setset{C'}$, 
  $||\setset{C} \cap \setset{C'}||\leq |\parents(\setset{C'})|$,
   (ii) $\text{rank}(\Sigma_{\setset{C}\cup \setset{X}, \setset{X}\cup\setset{X}_\graph \backslash \setset{C}\backslash\meassureddes(\setset{C})})=k$,
  then there exists an atomic cover
  $\set{V}=\set{L}\cup\set{X}$ in $\graph$,
  with $\set{X}=\cup_{\set{X'} \in \setset{X}} \set{X'}$, $|\set{L}|=k-|\set{X}|$,
  and $\cup_{\set{C} \in \setset{C}} \set{C}\subseteq \purechildren(\set{V})$.
    \label{thm:URD}
\end{theorem}
\vspace{0.8mm}

Basically, Theorem~\ref{thm:URD} says that in certain conditions, we can build a unique connection between rank deficiency and atomic covers.
Therefore, we can identify atomic covers by first ensuring the required conditions and then searching for combinations of $\mathcal{C}$ and $\mathcal{X}$ that induce rank deficiency.
Next we briefly introduce how to design an algorithm that strives to fulfill these conditions, in order to cash out the rank-deficient property for identifying atomic covers in a graph, and consequently, identifying the whole causal structure over both latent and observed variables. More details can be found in our manuscript~\citep{dong2023versatile}.

\

\textbf{Rank-based Latent Causal Discovery Algorithm.}
Our rank-based latent causal discovery consists of three phases.

\textbf{Phase 1: Finding CI skeleton.}
In this phase we just follow the classical PC algorithm \citep{spirtes2000causation} to find the CI skeleton by using CI tests. Then we will rely on the information from the CI skeleton to portion the CI skeleton into subgraphs that might contain latent variables. Then we take these subgraphs as input to Phase 2.

\begin{figure}[t]
   \begin{subfigure}[b]{0.22\textwidth}
     \includegraphics[width=\textwidth]{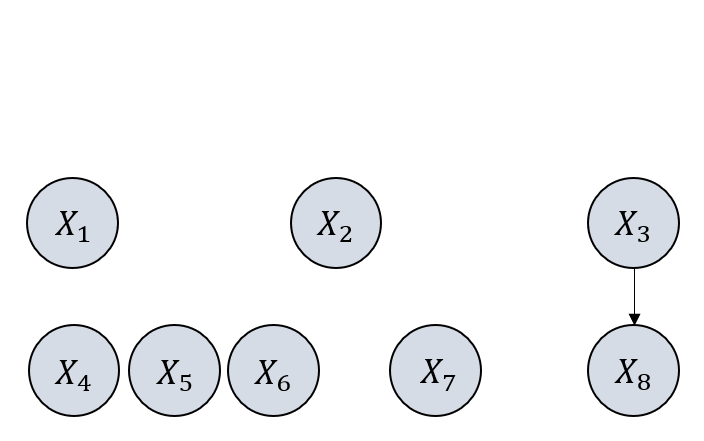}
     \caption{\scriptsize{Take $\setset{X}=\{\{\node{X_3}\}\}$ and \\$\setset{C}=\{\{\node{X_8}\}\}$.}}
 \end{subfigure}
 \hfill
 \begin{subfigure}[b]{0.22\textwidth}
   \includegraphics[width=\textwidth]{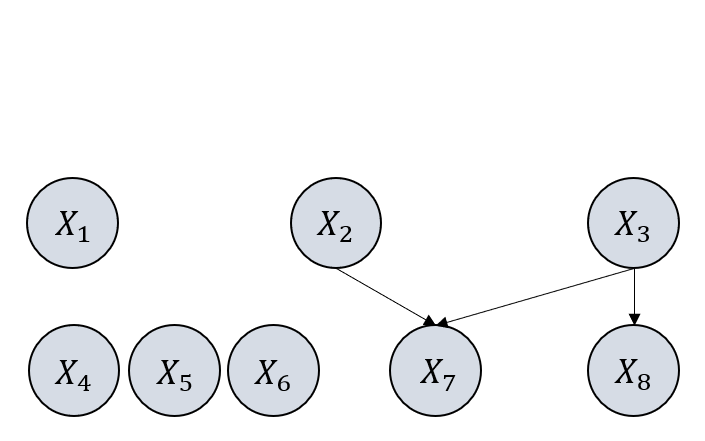}
   \caption{\scriptsize{Take $\setset{X}=\{\{\node{X_2},\node{X_3}\}\}$ and $\setset{C}=\{\{\node{X_7}\},\{\node{X_8}\}\}$.}}
 \end{subfigure}
 \hfill
 \begin{subfigure}[b]{0.22\textwidth}
   \includegraphics[width=\textwidth]{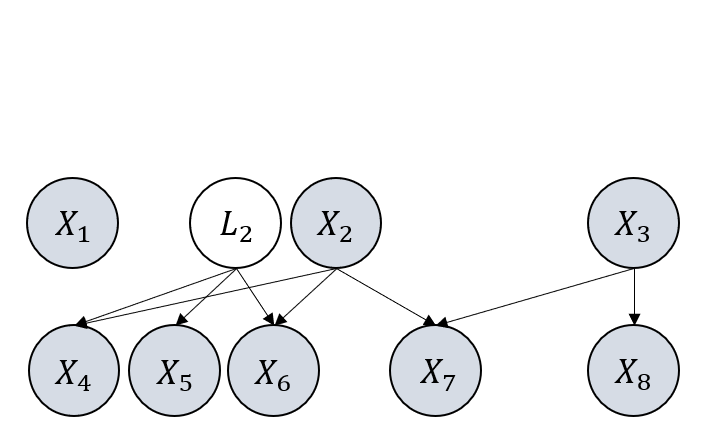}
   \caption{\scriptsize{Take $\setset{X}=\{\{\node{X_2}\}\}$ and \\$\setset{C}=\{\{\node{X_4}\},\{\node{X_5}\},\{\node{X_6}\}\}$.}}
 \end{subfigure}
 \hfill
 \begin{subfigure}[b]{0.22\textwidth}
   \includegraphics[width=\textwidth]{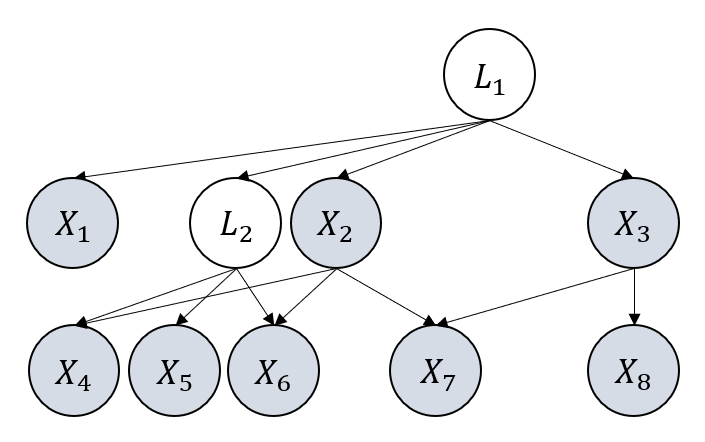}
   \caption{\scriptsize{Take $\setset{X}=\{\}$ and $\setset{C}=\{\{\node{X_1}\},\{\node{L_2},\node{X_2}\},\{\node{X_3}\}\}$}}
 \end{subfigure}
 \caption{\small An illustrative example of the key process of our rank-based latent causal discovery algorithm.}
   \label{fig:phase2}
 \end{figure}

\textbf{Phase 2: Finding causal clusters.} In this phase, for each sub-structure that might contain latent variables, we make use of  Theorem~\ref{thm:URD} to find every atomic covers in the graph. 
Thus we are interested in an effective search procedure to find combinations of 
sets of covers $\setset{C}$ and $\setset{X}$ (as defined in
Theorem~\ref{thm:URD}), where rank deficiency holds.
Meanwhile, 
our procedure should fulfill all the conditions required in Theorem~\ref{thm:URD} as much as possible in order to make the found rank deficiency uniquely indicate an atomic cover in $\graph$.

Phase 2 is the key procedure that identifies atomic covers and thus latent variables. Below we give an example with Figure~\ref{fig:phase2} to illustrate the procedure.

\begin{example}
Consider the graph in Figure~\ref{fig:phase2}.
  We start with finding atomic covers with $k=1$,
  and we can find that $\{\node{X_3}\}$ is a parent of $\{\node{X_8}\}$,
  as in Figure~\ref{fig:phase2}(a).
  At this point, no more $k=1$ clusters can be found, so next we search for $k=2$ clusters.
  Then to identify collider $\{\node{X_7}\}$, we only need to consider $\{\node{X_7}\}$ and $\{\node{X_8}\}$ together as the children of $\{\node{X_2},\node{X_3}\}$.
  %
  %
  After finding such a relationship, we arrive at Figure~\ref{fig:phase2}(b),
  and from now on the collider $\{\node{X_7}\}$ will not induce unfavorable rank deficiency anymore (as it is recorded).
  The next step is to find the relation of  $\{\node{X_4}\}$,$\{\node{X_5}\}$,$\{\node{X_6}\}$ with $\{\node{L_2},\node{X_2}\}$, 
   by
  taking $\setset{X}=\{\{\node{X_2}\}\}$ and $\setset{C}=\{\{\node{X_4}\},\{\node{X_5}\}\}$
  or $\{\{\node{X_4}\},\{\node{X_6}\}\}$ or $\{\{\node{X_5}\},\{\node{X_6}\}\}$,
  and thus conclude $\{\{\node{X_4}\},\{\node{X_5}\},\{\node{X_6}\}\}$ as the pure children of $\{\node{L_2},\node{X_2}\}$,
   as in Fig~\ref{fig:phase2}(c).
   Finally we 
    are able to find the relationship of $\{\node{X_1}\}$,$\{\node{L_2},\node{X_2}\}$, $\{\node{X_3}\}$ with $\{\node{L_1}\}$,
  by taking $\setset{X}=\{\}$ and $\setset{C}$ as $\{\{\node{X_1}\},\{\node{L_2},\node{X_2}\}\}$,
  $\{\{\node{X_1}\},\{\node{X_3}\}\}$, or $\{\{\node{L_2},\node{X_2}\},\{\node{X_3}\}\}$, as in Figure~\ref{fig:phase2}(d).
\end{example}

\textbf{Phase 3: Refining causal clusters.} 
In Phase 2, we strive to fulfill all required conditions such that we can correctly identify causal clusters and related structures.
However,
 there still exist some rare cases where our search cannot ensure the requirement (i) in Theorem~\ref{thm:URD}.
In this situation, Phase 2 might produce a big cluster in the resulting $\graphp$ that should be split into smaller ones.
Fortunately, the incorrect cluster will not do harm to the identification of other substructures in the graph, and thus in this phase we  characterize and refine the incorrect ones.

\

\noindent\textbf{Identifiability Theory.} 
The overall identifiability theory of the proposed causal discovery algorithm is summarized in the following theorem.

\begin{condition}[Graphical condition  on  colliders for identifiability] 
In a latent graph $\graph$, if there exists a set of  variables $\set{C}$
such that every variable in $\set{C}$ is a collider of two atomic covers $\set{V_1}$, $\set{V_2}$, and denote by $\set{A}$ the minimal set of variables that d-separates $\set{V_1}$ from $\set{V_2}$, then we have 
$|\set{C}| + |\set{A}| \geq |\set{V_1}|+|\set{V_2}|$.
\label{cond:vstructure}
\end{condition}

\begin{theorem}[Identifiability of Latent Causal Graphs]
Suppose $\graph$ is a DAG associated with a Linear Latent Causal Model
that satisfies Condition~\ref{cond:basic}
and Condition~\ref{cond:vstructure}.
Our rank-based latent causal discovery algorithm can asymptotically identify the Markov equivalence of $\mathcal{O}_{\text{min}}(\mathcal{O}_s(\graph))$.
\label{theorem:identifiability}
\end{theorem}

The graph operators $\mathcal{O}_{\text{min}}$ and $\mathcal{O}_s$ in the above theorem are defined following \citep{huang2022latent}; they are graphical operators that will not change the observational rank constraints.
Roughly speaking Theorem~\ref{theorem:identifiability} is saying that, if the required conditions are satisfied, our method can find a compact graphical representation of the underlying causal structure, with only some uncertainty about the direction of some edges. 

\subsection{Determining All Directions and Estimating Causal Coefficients}

Our rank-based latent causal discovery algorithm is able to output the Markove equivalence class of $\mathcal{O}_{\text{min}} (\mathcal{O}_{s}(\graph))$. The skeleton of the graph is clear, and yet some directions of edges cannot be decided.
Therefore, our next step is to determine all the rest undecided directions.

\textbf{Determining Causal Directions.}
Here we aim at making use of Generalized Independent Noise (GIN) condition \citep{xie2020generalized} to decide the directions, by assuming the noise terms $\epsilon_{\node{V}_{i}}$ in Definition~\ref{definition:lcm} are non-gaussian.
The GIN condition is given as follows.

\begin{condition} [GIN condition \citep{xie2020generalized}]
   Assume an LLCM with non-gaussian additive noise. Let $\set{Z}$ and $\set{Y}$ be two observed random vectors. Define the surrogate for $\set{Y}$ relative to $\set{Z}$ as $E_{\set{Y}||\set{Z}}:=\omega^T\set{Y}$,
   where $\omega$ satisfies $\omega^T\setsetset{E}[\set{Y}\set{Z}^T]=0$ and $\omega\neq0$. We say that $(\set{Z},\set{Y})$ follows the GIN condition if and only if $E_{\set{Y}||\set{Z}}$ is independent from $\set{Z}$.
   \end{condition}
It has been shown that, by making use of GIN conditions in specific ways, we can decide the directions of edges between latent variables \cite{xie2020generalized}, summarized as follows. 

\begin{theorem}[Deciding Directions by GIN \citep{xie2020generalized}]
 Assume an LLCM with non-gaussian additive noise. Let $\set{V_p}$ and $\set{V_q}$ be two atomic covers and there is no latent confounder behind $\set{V_p}$ and $\set{V_q}$, and $\set{V_p}\cap\set{V_q}=\emptyset$. Further suppose that $\set{V_p}$ has $2|\set{V_p}|$ pure children $\{\node{P_i}\}_1^{2|\set{V_p}|}$,
    and $\set{V_q}$ has $2|\set{V_q}|$ pure children $\{\node{Q_i}\}_1^{2|\set{V_q}|}$.
    If ($\set{Z}=\{\node{P_i}\}_1^{|\set{V_p}|}$, $\set{Y}=\{\node{P_i}\}_{|\set{V_p}|+1}^{2|\set{V_p}|}\cup\{\node{Q_i}\}_1^{|\set{V_q}|}$) follows the GIN condition, then $\set{V_p}\rightarrow\set{V_q}$.
    \label{thm:direction}
\end{theorem}
Theorem~\ref{thm:direction} is saying that even though we cannot directly observe latent variables, 
we can still decide the directions of edges between these latent variables by using observed variables as surrogates. Therefore, by first conducting our rank-based latent causal discovery algorithm and then checking directions following Theorem~\ref{thm:direction},
we are able to discover the skeleton together with the directions of the underlying graph.

\textbf{Estimating Causal Coefficients.}
Here we are interested in estimating the causal coefficients of each edge, given the causal structure and observational data. We provide our novel solution as follows. By the matrix form of Definition~\ref{definition:lcm}, i.e., 
$\set{V}=A\set{V}+\mathbf{\epsilon_{\set{V}}}$,
we have $\set{V}=(\set{I}-A)^{-1}\mathbf{\epsilon_{\set{V}}}$, and 
$\Sigma_{\set{V}}=(\set{I}-A)^{-1}\Psi(\set{I}-A)^{-T}$, where $\Psi$ is a diagonal matrix representing the covariance of $\epsilon_{V}$.
We can take the submatrices from $\Sigma_{\set{V}}$, and thus we have $\Sigma_{\set{X}}$ and $\Sigma_{\set{L}}$ parameterized by $A$ and $\Psi$.
Similar to the typical factor analysis setting \citep{gorsuch2014factor},
we assume that $\epsilon_{V}$ are all gaussian and thus $\set{X}$ are jointly gaussian, and we find the causal coefficients $A$ by maximizing the log-likelihood of observational data:
\begin{align}
&\arg\max_{A,\Psi} -\frac{n|\set{X}|}{2}\log 2\pi - \frac{n}{2}\log |\Sigma_{\set{X}}|
-\frac{n}{2}\text{tr}((\Sigma_{\set{X}})^{-1}\hat{\Sigma}_{\set{X}}),\\ 
&~\text{s.t.,}~\text{diagonal entries of }~\Sigma_{\set{L}}~\text{are all 1,}
    \vspace{1mm}
\end{align}
where $n$ is the number of i.i.d. observations of $\set{X}$, $\text{tr}(\dot)$ returns the trace of a matrix, and  $\hat{\Sigma}_{\set{X}}$ is the sample covariance estimated from data. We solve the optimization by gradient descent. We note that it is a non-convex optimization and thus we shall rely on multiple random initializations to approach the global optimum.

\subsection{The Whole Procedure and Potential Usage}

We will apply the proposed rank-based latent causal discovery method to SP100 stock price data,
which suffers from the three demos - time resolution mismatch, non-stationarity, and the existence of latent variables.

To deal with time resolution, mismatch, we assume the underlying procedure follows Eq.~\ref{eq1} and we use the log return of each stock, formulated as $\text{return}_t=\log(\text{price}_t-\text{price}_{t-1})$. By doing so, our observation of return
can be taken as linearly aggregated data, as in Eq.~\ref{eq2}, and thus 
we can take our data as the i.i.d. case to apply instantaneous causal discovery methods.

Regarding the non-stationarity, we will apply Baysian change point detection \cite{fearnhead2006exact,adams2007bayesian,xuan2007modeling} to identify the changing points in the time series data.
Once we detect the change point, we partion the time series into segments, and thus within each segments the time series can be taken as stationary.

As for the existence of latent variable, we apply our rank-based latent causal discovery method, together with GIN condition to identify the direction. We further apply our propose method to estimate all the causal coefficients.
Therefore, our final result include the causal graph, together with the coefficients of all the edges.

The potential usage of our result could be very broad. One significant usage would be estimating  causal effects using only observational data,
by making use of 
simulated interventions (\citep{pearl2009causality,lopez2022causal}).
This is a very promising direction and we recommend readers to refer to \cite{lopez2022causal} for how that can be achieved based on the knowledge of a causal model (which is the result we get) and observational distribution, especially in a financial context.

%% file: sec/5_experiments.tex
\section{Experiments}\label{sec:experiments}

In this section, we apply our proposed method to Sp100 stock price dataset and see how well we can address the three issues, i.e., time resolution, non-stationarity, and latent variables.

For the first issue, throughout the experiments we use the log daily/hourly return data, which can be written as the sum of the log minute return. Therefore, suppose the causal effects among stocks can be approximated as linear and happen instantaneously, according to~\cref{sec:time_resolution}, those instantaneous causal relations can still be recovered from the aggregated sum data. This can be verified by the reasonable results shown below. For the rest of this section, we focus on dealing with nonstationary relations and latent variables.

\subsection{Results on Constraint-based causal Discovery from heterogeneous/NOnstationary Data (CD-NOD)}

The scatter plot in~\cref{fig:scatter_plot} highlights the potential for significant shifts in causal relationships among stocks during the Covid-19 pandemic. To further investigate this and explore what information can obtained in a nonstationary market context, in this subsection we focus on ten representative stocks from the SP100. These stocks are across various sectors that might be influenced by the pandemic:

\begin{itemize}
    \item \textbf{Pharmaceuticals:} Pfizer (PFE) and Merck (MRK). These companies are directly involved in the development and distribution of Covid-19 treatments and vaccines.
    \item \textbf{Entertainment and Streaming:} Netflix (NFLX), Disney (DIS). The stay-at-home policies and social distancing measures increased demand for home entertainment.
    \item \textbf{Travel and Hospitality:} Boeing (BA), Booking Holdings (BKNG). Disney (DIS) may also be counted here as for their resorts. These companies experienced significant downturns due to travel restrictions and reduced consumer travel interest.
    \item \textbf{Consumer Services and Retail:} Walmart (WMT), Visa (V). These companies represent the consumer behavior and spending. Walmart saw changes in demand for essentials, while Visa saw overall consumer spending and economic activities.
    \item \textbf{Technology:} Apple (AAPL). Represents the overall technology sector.
    \item \textbf{Energy:} Chevron (CVX). May be impacted by the global oil demand and prices.
\end{itemize}

\begin{figure}[t]
\center
\includegraphics[width=0.4\textwidth]{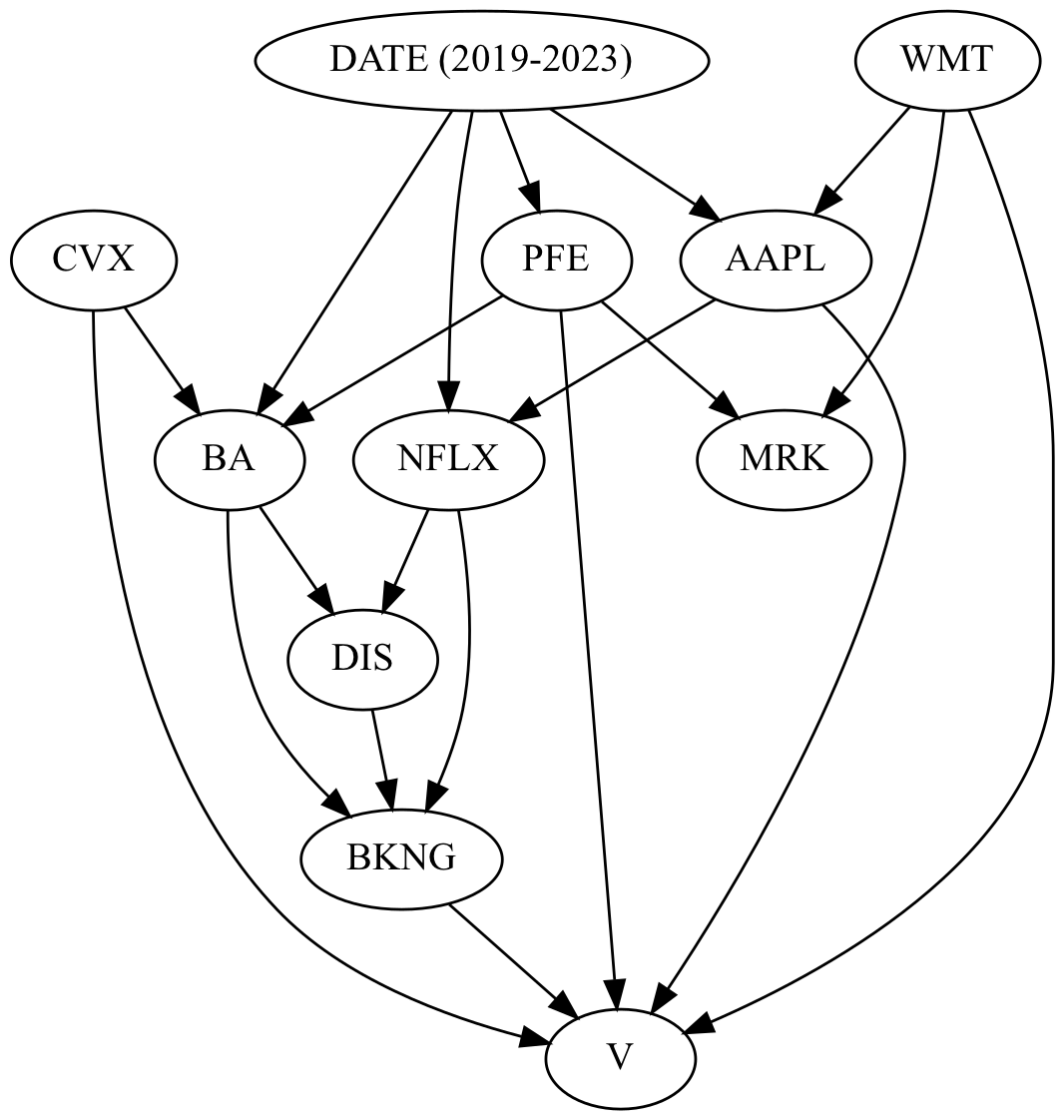}
\caption{CD-NOD discovered causal graph on the ten selected stocks from SP100 using data from 2019 to 2023. The top ``DATE'' is the date index serving as a surrogate for nonstationarity in the dataset. Variables that are direct child of ``DATE'' are those whose causal generating mechanisms (i.e., the conditional distribution of this variable give its other stock parental variables) are subject to changes over time.}
\label{fig:cdnod_graph_results}
\end{figure}

\begin{figure}[t]
\center
\includegraphics[width=0.7\textwidth]{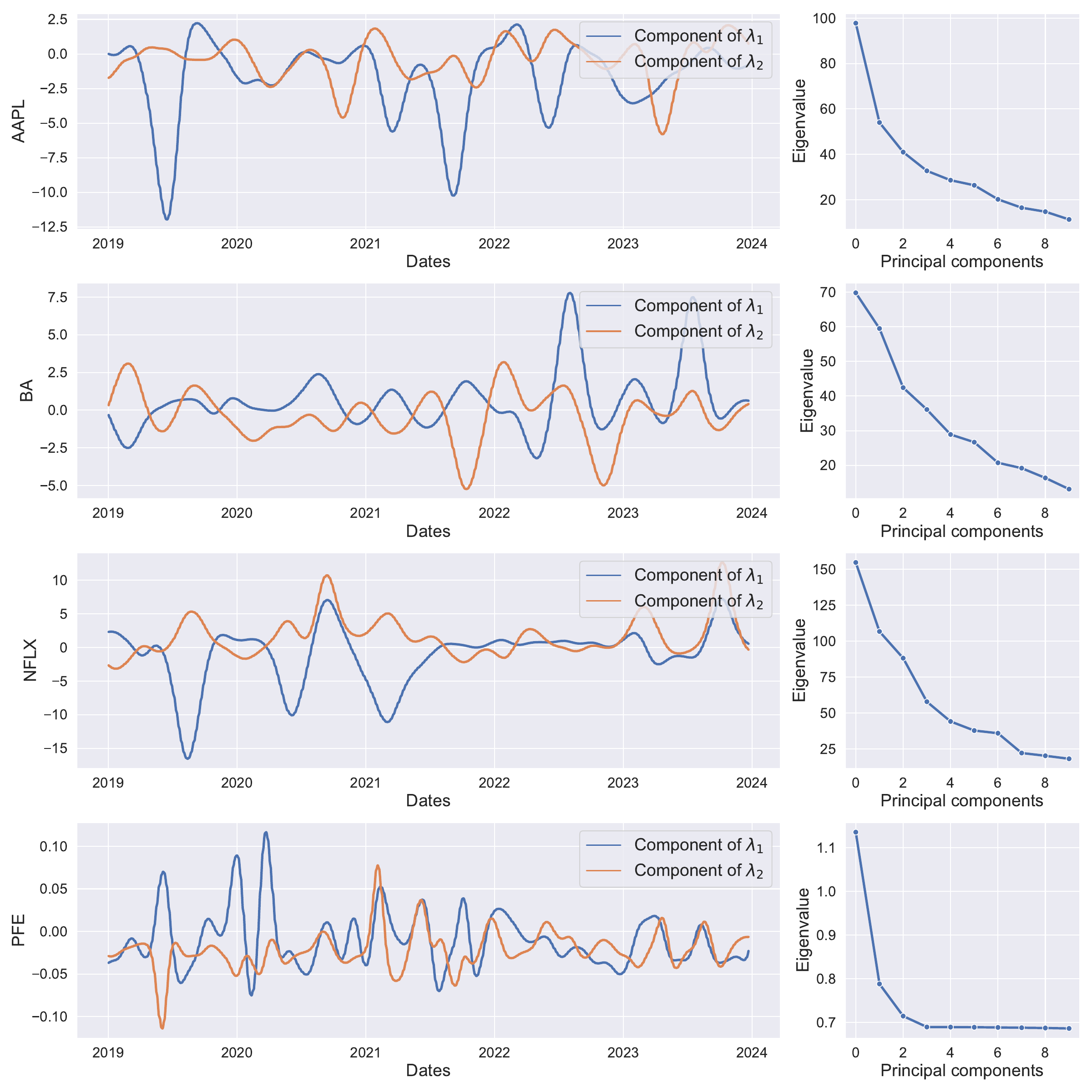}
\caption{Visualization of estimated driving forces of changing causal modules using CD-NOD's Phase III. Each row indicates one stock whose causal module changes over time. The left panel shows the so-called ``driving force'', i.e., the top-2 principle nonstationary components recovered by Kernel Nonstationary Visualization (KNV). Note that the lines primarily indicate ``change points'' in trends; they do not represent actual prices, nor do they carry a direct physical interpretation on the y-axis. The right panel shows the largest ten eigenvalues of the corresponding Gram matrix.}
\label{fig:cdnod_change_results}
\end{figure}

The CD-NOD analysis was conducted on these ten selected stocks, utilizing their log daily return data spanning from 2019 to 2023. The discovered causal graph and the estimated ``driving forces'' of changing causal modules are shown in~\cref{fig:cdnod_graph_results} and~\cref{fig:cdnod_change_results}, respectively.

The discovered causal graph indicates four stocks whose causal generating mechanisms are subject to changes over time: Pfizer (PFE), Boeing (BA), Apple (AAPL), and Netflix (NFLX). The change on PFE may be because that the new drug (and especially Covid-19 vaccines) researches and approvals or health crises can suddenly alter the landscape. BA could be subject to changes due to the significant fluctuations in air travel throughout these recent years. NFLX saw the change in the rapidly evolving media streaming space, and this change may be especially significant during the Covid-19 times during to high demand. Similarly, AAPL operates in a highly dynamic tech market where where innovation and competition naturally shift over time.

To specifically demonstrate how the causal modules of these four variables change over time, CD-NOD's Phase III is performed and the result is shown in~\cref{fig:cdnod_change_results}. Several of the identified changes align with known events and can be reasonably explained. For instance, the notable changes in Pfizer (PFE) during the early stages of 2020 and the middle of 2021 coincide with the emergence of COVID-19 and subsequent vaccine development milestones. The changes in Netflix (NFLX) (more specifically, the density $p$(NFLX|AAPL)) around late 2020 and early 2021 correspond to their increasing subscriber base. Other changes might not stem directly from the pandemic but could be influenced by various factors over time.

Across the whole causal graph, most of the edges (direct causal effects) seem also reasonable: The effect from PFE to BA, as echoed in~\cref{fig:scatter_plot}, may not typically be evident but make sense in the context of the Covid-19 pandemic, where the development of vaccines likely bolstered traveler confidence, impacting the aerospace sector. The direct relationships among Boeing (BA), Disney (DIS), and Booking Holdings (BKNG) may reveal the causal effects within the travel industry. Visa (V) being a joint effect of multiple stocks across different sectors demonstrates how consumer spending patterns, influenced by the broader economic environment, can reflect on payment processing companies. Some other edges lack a straightforward interpretation though, such as edges from Walmart (WMT) to Apple (AAPL) and Merck (MRK).

The limitations of the CD-NOD results presented here lie in that the analysis is restricted to just ten stocks, not the entire market. This constraint arises from the necessity to employ general, nonparametric conditional independence tests (e.g., the kernel-based conditional independence (KCI) test~\citep{zhang2012kernel}) to detect the nonlinear relationships associated with the surrogate variable (time index): while it's reasonable to assume linearity among stock variables, the nonstationarity of causal modules, represented by the relationship of other variables on time index, tends to be inherently nonlinear. However, KCI-test has a cubic time complexity relative to sample size, which therefore limits the number of variables considered due to time constraints. To address this issue, we move forward from another perspective: with causal discovery that incorporates latent variables (note that aligned with the surrogate time index here, the nonstationarity can also be represented by a latent variable).

\begin{figure}[t]
\center
 \begin{subfigure}[b]{0.7\textwidth}
   \includegraphics[width=\textwidth]{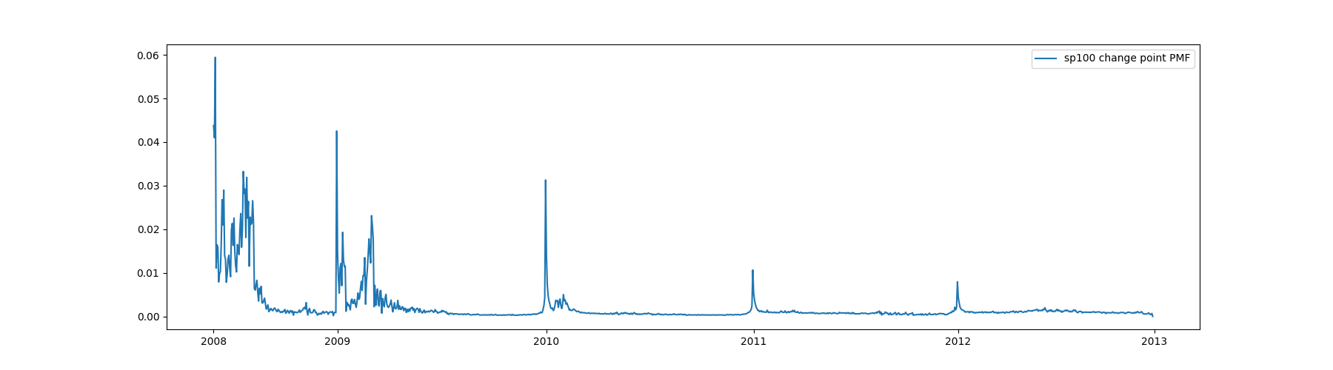}
   \caption{Change point detection for SP100 from 2008 to 2013. The detected change point in 2008 can be explained as  Global financial crisis and subprime mortgage crisis starting from 2007. The change point in 2009 can be explained as European debt crisis starting from 2009.}
 \end{subfigure}
 \begin{subfigure}[b]{0.7\textwidth}
   \includegraphics[width=\textwidth]{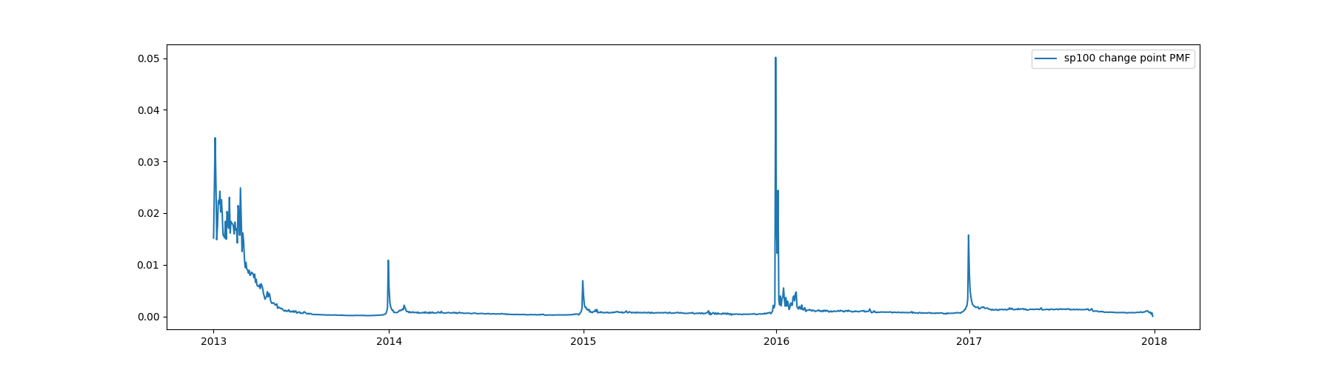}
   \caption{Change point detection for SP100 from 2013 to 2018.
The detected change point in 2013 could relate to Cypriot financial crisis in 2013.
   The detected change point in 2016 could relate to Chinese stock market 
crash in 2016.}
 \end{subfigure}
 \begin{subfigure}[b]{0.7\textwidth}
   \includegraphics[width=\textwidth]{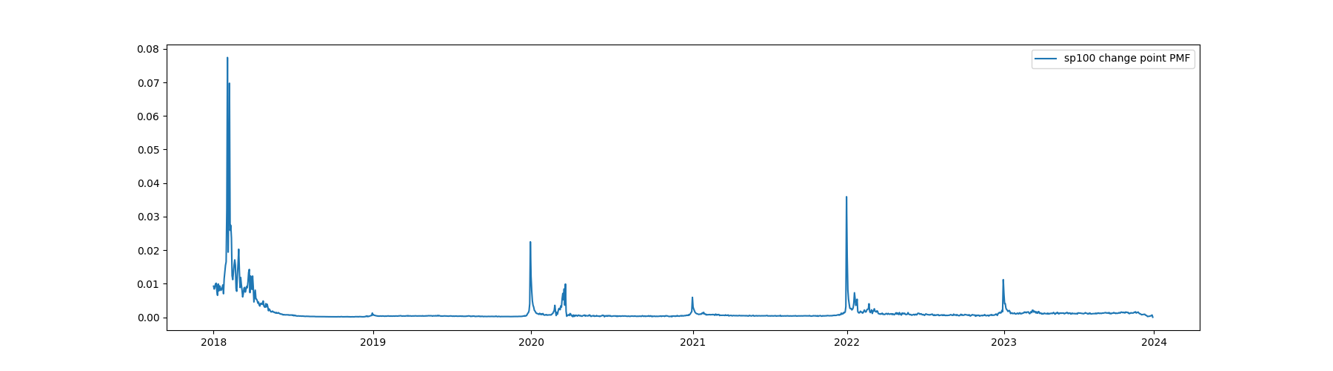}
   \caption{Change point detection for SP100 from 2018 to 2023. The detected change point in 2018 could relate to cryptocurrency meltdown. The change point in 2020 likely relates to Covid-19. The change point in 2022 could relate to the Russian-Ukraine war. }
 \end{subfigure}
    \caption{Baysian changing point detection result for SP100 from 2008 to 2023.}
    \label{fig:bcpd}
\end{figure}

\subsection{Baysian Change Point Detection}

To utilize our rank-based latent causal discovery method we need to assume that our data is stationary;
otherwise the nonstationary would induce rank deficiency constraints that cannot correctly reflect the underlying causal structure.
Therefore, we first employ Baysian change point detection \cite{fearnhead2006exact,adams2007bayesian,xuan2007modeling} to identify the changing points in the time series data.
We take the average log daily return over all Sp100 stocks and take that as input to the Baysian change point detection.

Our results are shown in Figure~\ref{fig:bcpd},
where most of the change points detected can relate to real-life financial crises or events.
Specifically, the change point in 2008 can be explained as Global financial crisis and subprime mortgage crisis starting from 2007. The change point in 2009 can be explained as European debt crisis starting from 2009.
The detected change point in 2016 could relate to Chinese stock market 
crash in 2016.
The change point in 2018 could relate to cryptocurrency meltdown. The change point in 2020 relates to Covid-19. The change point in 2022 could relate to the Russian-Ukraine war.

The detected change points in 2018 and 2020 are significant and are well related to real-life events. Thus we split the SP100 data using these two change points and have three time series: (i) SP100 from Jan 2017 to Dec 2017,  (ii) SP100 from Jan 2019 to Dec 2019, and (iii) SP100 from Jan 2021 to Dec 2021. We will refer to them as SP100-2017, SP100-2019, and SP100-2021 respectively in the following. As we do not detect any significant change point inside these three years, they can be considered as stationary time series.

In the next section, we will show the causal discovery results using these three time series and compare the common part as well as the difference.

\begin{figure}[t]
\center
\begin{subfigure}[b]{1\textwidth}
   \centering
   \includegraphics[width=0.36\textwidth]{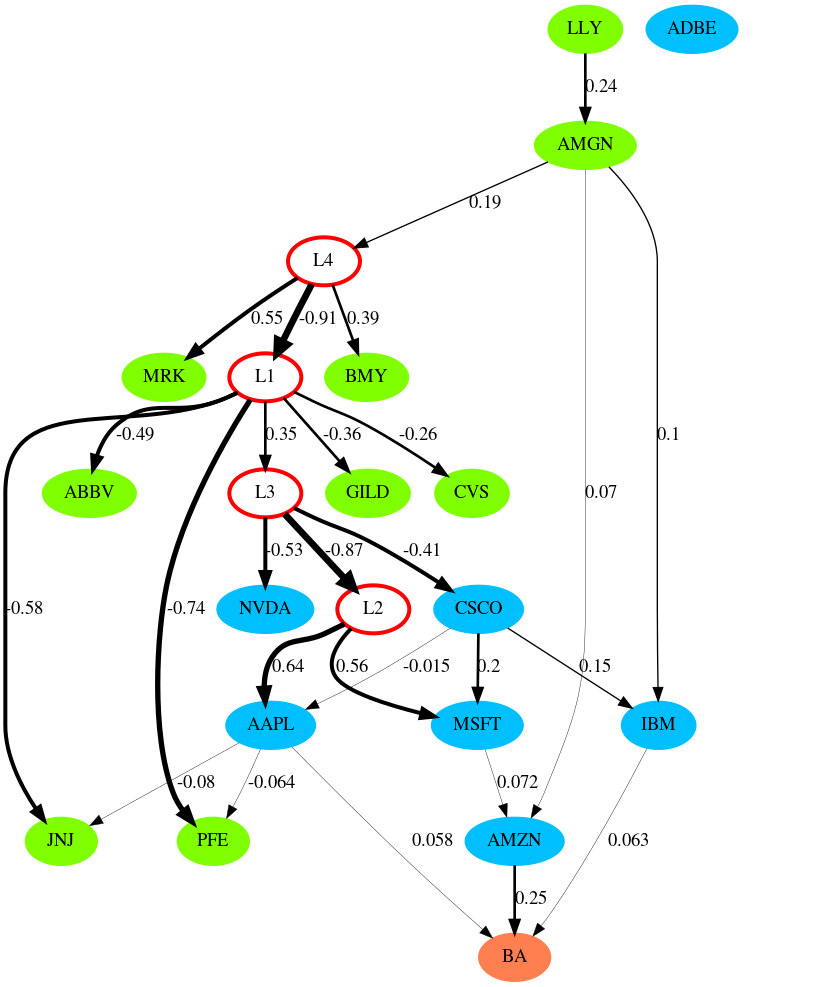}
   \caption{Causal discovery result with causal coefficients for Sp100-2017.}
 \end{subfigure}
 \begin{subfigure}[b]{1\textwidth}
   \centering
   \includegraphics[width=0.5\textwidth]{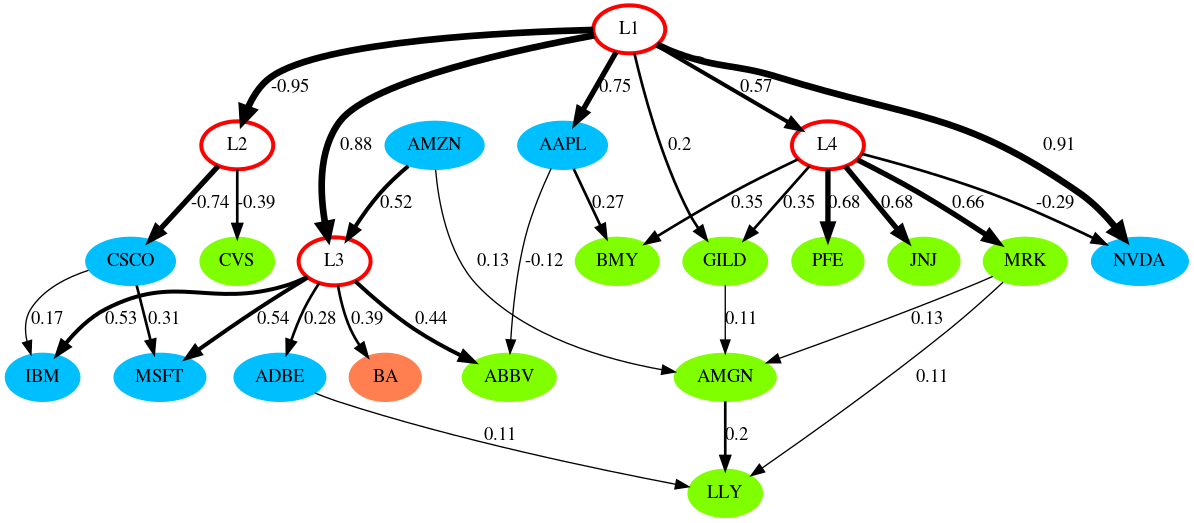}
   \caption{Causal discovery result with causal coefficients for Sp100-2019.}
 \end{subfigure}
  \begin{subfigure}[b]{1\textwidth}
   \centering
   \includegraphics[width=0.5\textwidth]{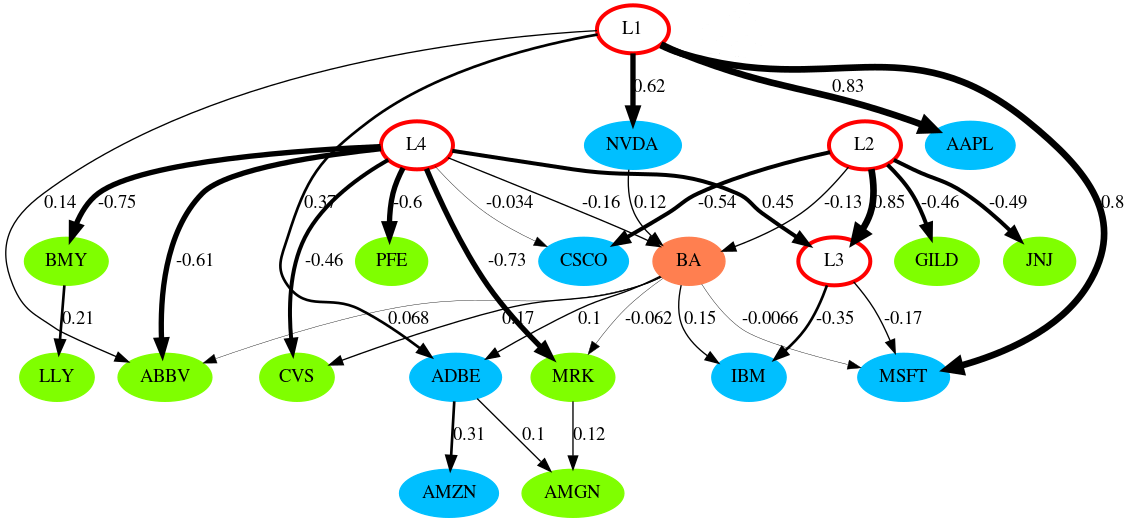}
   \caption{Causal discovery result with causal coefficients for Sp100-2021.}
 \end{subfigure}
     \caption{Discovered causal graph for Sp100-2017, Sp100-2019, and Sp100-2021, respectively.
    The three graphs have many things in common, and yet a big difference between 2019 and 2021 is that travel-related stocks are very causally related to pharmaceutical stocks (or pharm-related latent variables) after Covid-19, while before Covid-19 it is not the case.
    }
    \label{fig:rank_three_years}
\end{figure}

\subsection{Rank-based Latent Causal Discovery with Causal Coefficients}

In this section, we apply our rank-based latent causal discovery method to the Sp100 time series data. 
In addition, we further apply the GIN condition to determine all the directions and estimate the coefficients of all the edges.
 For the change point detection we used log daily return but for rank-based latent causal discovery we use log houlry return such that we can have more datapoints within each time series. All the variables are normalized to have zero mean and unit variance.

Our first experiment is to show how the change point, e.g., Covid-19, influences the underlying graph. To this end we apply our method to the three
time series data Sp100-2017, Sp100-2019, and Sp100-2021 respectively.
To better compare the common parts and the difference, we only pick a subset of the Sp100 stocks.
They are travel-related stocks (BA) colored in orange, pharmaceutical stocks (BMY, PFE, JNJ, MRK, GILD, ABBV, CVS, LLY, AMGN) colored in green, and info-tech stocks (ADBE, IBM, CSCO, NVDA, MSFT, AAPL, AMZN) colored in blue.

The result is shown in Figure~\ref{fig:rank_three_years}.
We can see that these three graphs have things in common. The most significant pattern is that all the stocks are basically clustered by the type of the company. For example, in Sp100-2017,
there are basically two latent variables influencing all the pharmaceutical companies, and the same logic also applies to Sp100-2019 and Sp100-2021.
The difference among the three graphs is even more interesting. One prominent difference is that 
for both Sp100-2017 and Sp100-2019,
the BA, which is related to travelling, 
is more causally related to tech companies.
Specifically, in 2017, BA is caused by AMZN, AAPL, and IBM and in 2019, BA is caused by L3, which is the parent of tech companies MSFT, IBM, and ADBE.
In contrast, in 2021, BA is influenced by two latent variables, L2 and L4, as shown in (c) in Figure~\ref{fig:rank_three_years},
where L2 and L4 are the parents of many pharmaceutical companies.
This can be well explained by the change point caused by Covid-19. Once 
a new vaccination is approved or announced to be successful, the confidence in both the travel section and pharmaceutical industry increases at the same time; on the other hand, if there are setbacks in vaccine development or distribution, airline stocks may suffer due to anticipated continued travel restrictions.

\begin{figure}[t]
\center
\includegraphics[width=0.75\textwidth]{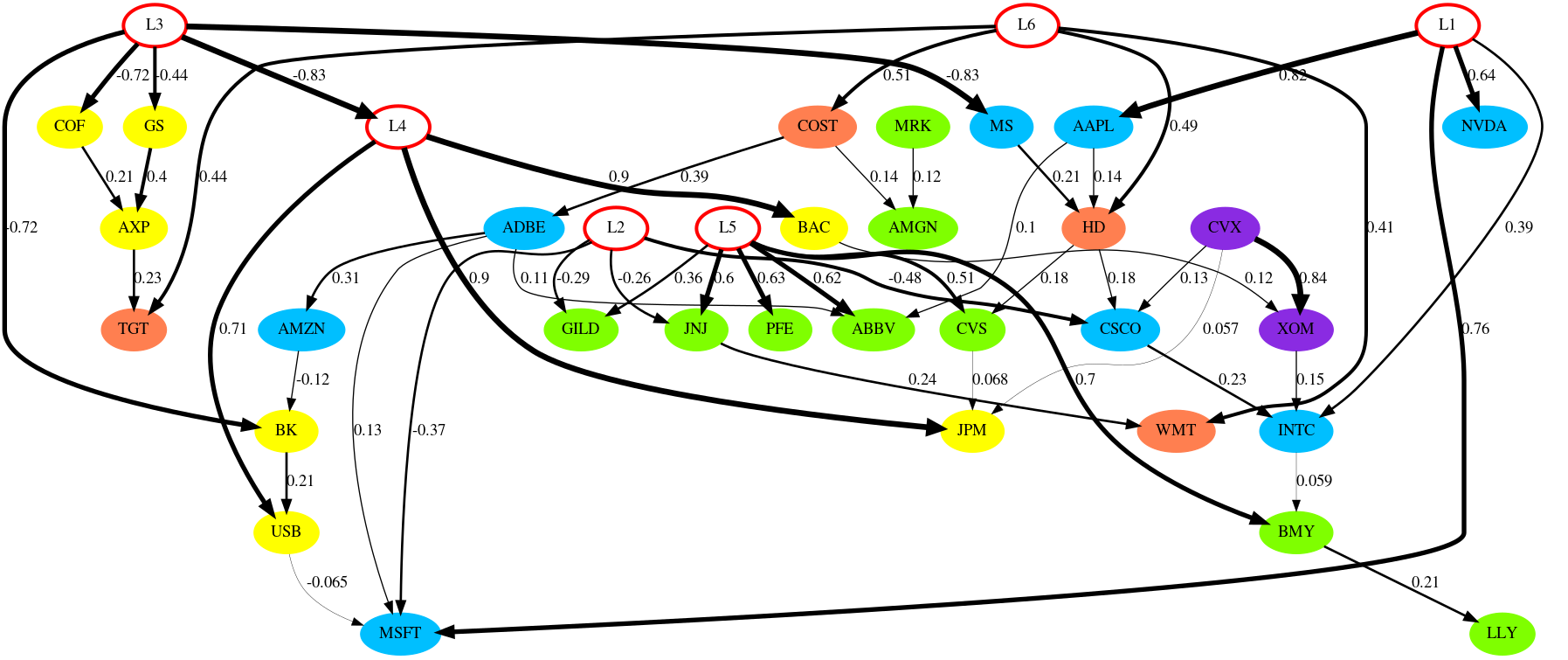}
   \caption{Causal discovery result with causal coefficients for Sp100-2021, with stocks from five sections. They are pharmaceutical stocks (BMY, PFE, JNJ, MRK, GILD, ABBV, CVS, LLY, AMGN) colored in green, and info-tech stocks (ADBE, IBM, CSCO, NVDA, MSFT, AAPL, AMZN, INTC) colored in blue,
energy companies (CVX, XOM) colored in purple, finance companies (COF, GS, AXP, BK, USB, BAC, JPM) colored in yellow, and consumer-related companies (TGT, COST, WMT, HD) colored in orange.}
    \label{fig:rank_final}
\end{figure}

We further apply our method to Sp100-2021 with even more stocks from more sections.
Specifically, they are pharmaceutical stocks (BMY, PFE, JNJ, MRK, GILD, ABBV, CVS, LLY, AMGN) colored in green, and info-tech stocks (ADBE, IBM, CSCO, NVDA, MSFT, AAPL, AMZN, INTC) colored in blue,
energy companies (CVX, XOM) colored in purple, finance companies (COF, GS, AXP, BK, USB, BAC, JPM) colored in yellow, and consumer-related companies (TGT, COST, WMT, HD) colored in orange.

The discovered causal graph with all the causal coefficients are shown in Figure~\ref{fig:rank_final}.
As expected, all the stocks from the same section are well clustered. For example, all consumer goods fall under single Latent L6 with relatively high coefficients.
All financial companies are under single Latent L3 with one sub group under another representative L4 which again is a child of L3. This also implies that BAC, USB, JPM
share something in common that is not shared by other financial stocks.
As for the energy companies XOM and CVX,
we observe that they are highly related with a high coefficient 0.84. At the same time they are also causally related to some other companies including CSCO, JPM, and INTC, with relatively small strength of coefficient.

Based on our causal discovery result, we can unveil more interesting findings, such as estimating causal effect by simulated intervention, and validating results by further comparing to real interventions, as suggested in \cite{lopez2022causal}. As it is beyond of the scope of this paper, we will leave it for future work.

%% file: sec/7_conclusion_discussion.tex
\section{Conclusion and Discussion}
In this paper we discuss the demons in causality in finance: time resolution, nonstationarity, and latent factors. We follow a systematic causal perspective to look into these three issues and provided rigorous methods to deal with them. Our preliminary findings reveal compelling and novel insights as well as a foundation for further exploration.

%% file: main.bbl
\begin{thebibliography}{52}
\providecommand{\natexlab}[1]{#1}
\providecommand{\url}[1]{\texttt{#1}}
\expandafter\ifx\csname urlstyle\endcsname\relax
  \providecommand{\doi}[1]{doi: #1}\else
  \providecommand{\doi}{doi: \begingroup \urlstyle{rm}\Url}\fi

\bibitem[Adams and MacKay(2007)]{adams2007bayesian}
Ryan~Prescott Adams and David~JC MacKay.
\newblock Bayesian online changepoint detection.
\newblock \emph{arXiv preprint arXiv:0710.3742}, 2007.

\bibitem[Akbari et~al.(2021)Akbari, Mokhtarian, Ghassami, and Kiyavash]{akbari2021recursive}
Sina Akbari, Ehsan Mokhtarian, AmirEmad Ghassami, and Negar Kiyavash.
\newblock Recursive causal structure learning in the presence of latent variables and selection bias.
\newblock \emph{Advances in Neural Information Processing Systems}, 34:\penalty0 10119--10130, 2021.

\bibitem[Bach and Maloof(2008)]{bach2008paired}
Stephen~H Bach and Marcus~A Maloof.
\newblock Paired learners for concept drift.
\newblock In \emph{2008 Eighth IEEE International Conference on Data Mining}, pages 23--32. IEEE, 2008.

\bibitem[Breitung and Swanson(2002)]{breitung2002temporal}
J{\"o}rg Breitung and Norman~R Swanson.
\newblock Temporal aggregation and spurious instantaneous causality in multiple time series models.
\newblock \emph{Journal of Time Series Analysis}, 23\penalty0 (6):\penalty0 651--665, 2002.

\bibitem[Chickering(2002)]{chickering2002optimal}
David~Maxwell Chickering.
\newblock Optimal structure identification with greedy search.
\newblock \emph{Journal of machine learning research}, 3\penalty0 (Nov):\penalty0 507--554, 2002.

\bibitem[Colombo et~al.(2012)Colombo, Maathuis, Kalisch, and Richardson]{colombo2012learning}
Diego Colombo, Marloes~H Maathuis, Markus Kalisch, and Thomas~S Richardson.
\newblock Learning high-dimensional directed acyclic graphs with latent and selection variables.
\newblock \emph{The Annals of Statistics}, pages 294--321, 2012.

\bibitem[Domingos and Hulten(2000)]{domingos2000mining}
Pedro Domingos and Geoff Hulten.
\newblock Mining high-speed data streams.
\newblock In \emph{Proceedings of the sixth ACM SIGKDD international conference on Knowledge discovery and data mining}, pages 71--80, 2000.

\bibitem[Dong et~al.(2023)Dong, Huang, Ng, Song, Zheng, Jin, Legaspi, Spirtes, and Zhang]{dong2023versatile}
Xinshuai Dong, Biwei Huang, Ignavier Ng, Xiangchen Song, Yujia Zheng, Songyao Jin, Roberto Legaspi, Peter Spirtes, and Kun Zhang.
\newblock A versatile causal discovery framework to allow causally-related hidden variables.
\newblock \emph{arXiv preprint arXiv:2312.11001}, 2023.

\bibitem[Fama and French(1996)]{fama1996multifactor}
Eugene~F Fama and Kenneth~R French.
\newblock Multifactor explanations of asset pricing anomalies.
\newblock \emph{The journal of finance}, 51\penalty0 (1):\penalty0 55--84, 1996.

\bibitem[Fama and French(2015)]{fama2015five}
Eugene~F Fama and Kenneth~R French.
\newblock A five-factor asset pricing model.
\newblock \emph{Journal of financial economics}, 116\penalty0 (1):\penalty0 1--22, 2015.

\bibitem[Fearnhead(2006)]{fearnhead2006exact}
Paul Fearnhead.
\newblock Exact and efficient bayesian inference for multiple changepoint problems.
\newblock \emph{Statistics and computing}, 16:\penalty0 203--213, 2006.

\bibitem[Fisher(1970)]{fisher1970correspondence}
Franklin~M Fisher.
\newblock A correspondence principle for simultaneous equation models.
\newblock \emph{Econometrica: Journal of the Econometric Society}, pages 73--92, 1970.

\bibitem[Gama et~al.(2014)Gama, {\v{Z}}liobait{\.e}, Bifet, Pechenizkiy, and Bouchachia]{gama2014survey}
Jo{\~a}o Gama, Indr{\.e} {\v{Z}}liobait{\.e}, Albert Bifet, Mykola Pechenizkiy, and Abdelhamid Bouchachia.
\newblock A survey on concept drift adaptation.
\newblock \emph{ACM computing surveys (CSUR)}, 46\penalty0 (4):\penalty0 1--37, 2014.

\bibitem[Gomes et~al.(2017)Gomes, Bifet, Read, Barddal, Enembreck, Pfharinger, Holmes, and Abdessalem]{gomes2017adaptive}
Heitor~M Gomes, Albert Bifet, Jesse Read, Jean~Paul Barddal, Fabr{\'\i}cio Enembreck, Bernhard Pfharinger, Geoff Holmes, and Talel Abdessalem.
\newblock Adaptive random forests for evolving data stream classification.
\newblock \emph{Machine Learning}, 106:\penalty0 1469--1495, 2017.

\bibitem[Gong et~al.(2017)Gong, Zhang, Sch{\"o}lkopf, Glymour, and Tao]{gong2017causal}
Mingming Gong, Kun Zhang, Bernhard Sch{\"o}lkopf, Clark Glymour, and Dacheng Tao.
\newblock Causal discovery from temporally aggregated time series.
\newblock In \emph{Uncertainty in artificial intelligence: proceedings of the... conference. Conference on Uncertainty in Artificial Intelligence}, volume 2017. NIH Public Access, 2017.

\bibitem[Gorsuch(2014)]{gorsuch2014factor}
Richard~L Gorsuch.
\newblock \emph{Factor analysis: Classic edition}.
\newblock Routledge, 2014.

\bibitem[Granger(1969)]{granger1969investigating}
Clive~WJ Granger.
\newblock Investigating causal relations by econometric models and cross-spectral methods.
\newblock \emph{Econometrica: journal of the Econometric Society}, pages 424--438, 1969.

\bibitem[Harvey and Chung(2000)]{harvey2000estimating}
Andrew Harvey and Chia-Hui Chung.
\newblock Estimating the underlying change in unemployment in the uk.
\newblock \emph{Journal of the Royal Statistical Society: Series A (Statistics in Society)}, 163\penalty0 (3):\penalty0 303--309, 2000.

\bibitem[Hausman and Woodward(1999)]{hausman1999independence}
Daniel~M Hausman and James Woodward.
\newblock Independence, invariance and the causal markov condition.
\newblock \emph{The British journal for the philosophy of science}, 50\penalty0 (4):\penalty0 521--583, 1999.

\bibitem[Hou et~al.(2015)Hou, Xue, and Zhang]{hou2015digesting}
Kewei Hou, Chen Xue, and Lu~Zhang.
\newblock Digesting anomalies: An investment approach.
\newblock \emph{The Review of Financial Studies}, 28\penalty0 (3):\penalty0 650--705, 2015.

\bibitem[Hou et~al.(2017)Hou, Xue, and Zhang]{hou2017comparison}
Kewei Hou, Chen Xue, and Lu~Zhang.
\newblock A comparison of new factor models.
\newblock \emph{Fisher college of business working paper}, \penalty0 (2015-03):\penalty0 05, 2017.

\bibitem[Hoyer et~al.(2008)Hoyer, Shimizu, Kerminen, and Palviainen]{hoyer2008estimation}
Patrik~O Hoyer, Shohei Shimizu, Antti~J Kerminen, and Markus Palviainen.
\newblock Estimation of causal effects using linear non-gaussian causal models with hidden variables.
\newblock \emph{International Journal of Approximate Reasoning}, 49\penalty0 (2):\penalty0 362--378, 2008.

\bibitem[Huang et~al.(2020)Huang, Zhang, Zhang, Ramsey, Sanchez-Romero, Glymour, and Sch{\"o}lkopf]{huang2020causal}
Biwei Huang, Kun Zhang, Jiji Zhang, Joseph Ramsey, Ruben Sanchez-Romero, Clark Glymour, and Bernhard Sch{\"o}lkopf.
\newblock Causal discovery from heterogeneous/nonstationary data.
\newblock \emph{The Journal of Machine Learning Research}, 21\penalty0 (1):\penalty0 3482--3534, 2020.

\bibitem[Huang et~al.(2022)Huang, Low, Xie, Glymour, and Zhang]{huang2022latent}
Biwei Huang, Charles Jia~Han Low, Feng Xie, Clark Glymour, and Kun Zhang.
\newblock Latent hierarchical causal structure discovery with rank constraints.
\newblock \emph{Advances in Neural Information Processing Systems}, 35:\penalty0 5549--5561, 2022.

\bibitem[Li et~al.(2022)Li, Yang, Liu, Xia, and Bian]{li2022ddg}
Wendi Li, Xiao Yang, Weiqing Liu, Yingce Xia, and Jiang Bian.
\newblock Ddg-da: Data distribution generation for predictable concept drift adaptation.
\newblock In \emph{Proceedings of the AAAI Conference on Artificial Intelligence}, volume~36, pages 4092--4100, 2022.

\bibitem[Lopez~de Prado(2022)]{lopez2022causal}
Marcos Lopez~de Prado.
\newblock Causal factor investing: Can factor investing become scientific?
\newblock \emph{Available at SSRN 4205613}, 2022.

\bibitem[Lu et~al.(2018)Lu, Liu, Dong, Gu, Gama, and Zhang]{lu2018learning}
Jie Lu, Anjin Liu, Fan Dong, Feng Gu, Joao Gama, and Guangquan Zhang.
\newblock Learning under concept drift: A review.
\newblock \emph{IEEE transactions on knowledge and data engineering}, 31\penalty0 (12):\penalty0 2346--2363, 2018.

\bibitem[Lu et~al.(2014)Lu, Zhang, and Lu]{lu2014concept}
Ning Lu, Guangquan Zhang, and Jie Lu.
\newblock Concept drift detection via competence models.
\newblock \emph{Artificial Intelligence}, 209:\penalty0 11--28, 2014.

\bibitem[Marcellino(1999)]{marcellino1999some}
Massimiliano Marcellino.
\newblock Some consequences of temporal aggregation in empirical analysis.
\newblock \emph{Journal of Business \& Economic Statistics}, pages 129--136, 1999.

\bibitem[Narang(2013)]{narang2013inside}
Rishi~K Narang.
\newblock \emph{Inside the black box: A simple guide to quantitative and high frequency trading}, volume 846.
\newblock John Wiley \& Sons, 2013.

\bibitem[Pearl(1988)]{pearl:88}
Judea Pearl.
\newblock \emph{Probabilistic {R}easoning in {I}ntelligent {S}ystems: {N}etworks of {P}lausible {I}nference}.
\newblock Morgan Kaufman Publishers, San Mateo, CA, 1988.

\bibitem[Pearl(2009)]{pearl2009causality}
Judea Pearl.
\newblock \emph{Causality}.
\newblock Cambridge university press, 2009.

\bibitem[Pearl et~al.(2000)]{pearl2000models}
Judea Pearl et~al.
\newblock Models, reasoning and inference.
\newblock \emph{Cambridge, UK: CambridgeUniversityPress}, 19\penalty0 (2):\penalty0 3, 2000.

\bibitem[Proietti(2006)]{proietti2006temporal}
Tommaso Proietti.
\newblock Temporal disaggregation by state space methods: Dynamic regression methods revisited.
\newblock \emph{The Econometrics Journal}, 9\penalty0 (3):\penalty0 357--372, 2006.

\bibitem[Rajaguru and Abeysinghe(2008)]{rajaguru2008temporal}
Gulasekaran Rajaguru and Tilak Abeysinghe.
\newblock Temporal aggregation, cointegration and causality inference.
\newblock \emph{Economics Letters}, 101\penalty0 (3):\penalty0 223--226, 2008.

\bibitem[Riva et~al.(2022)Riva, Bisi, Liotet, Sabbioni, Vittori, Pinciroli, Trapletti, and Restelli]{riva2022addressing}
Antonio Riva, Lorenzo Bisi, Pierre Liotet, Luca Sabbioni, Edoardo Vittori, Marco Pinciroli, Michele Trapletti, and Marcello Restelli.
\newblock Addressing non-stationarity in fx trading with online model selection of offline rl experts.
\newblock In \emph{Proceedings of the Third ACM International Conference on AI in Finance}, pages 394--402, 2022.

\bibitem[Runge et~al.(2019)Runge, Nowack, Kretschmer, Flaxman, and Sejdinovic]{runge2019detecting}
J~Runge, P~Nowack, M~Kretschmer, S~Flaxman, and D~Sejdinovic.
\newblock Detecting causal associations in large nonlinear time series datasets, sci. adv., 5, eaau4996, 2019.

\bibitem[Salehkaleybar et~al.(2020)Salehkaleybar, Ghassami, Kiyavash, and Zhang]{salehkaleybar2020learning}
Saber Salehkaleybar, AmirEmad Ghassami, Negar Kiyavash, and Kun Zhang.
\newblock Learning linear non-gaussian causal models in the presence of latent variables.
\newblock \emph{The Journal of Machine Learning Research}, 21\penalty0 (1):\penalty0 1436--1459, 2020.

\bibitem[Shimizu et~al.(2006)Shimizu, Hoyer, Hyv{\"a}rinen, Kerminen, and Jordan]{shimizu2006linear}
Shohei Shimizu, Patrik~O Hoyer, Aapo Hyv{\"a}rinen, Antti Kerminen, and Michael Jordan.
\newblock A linear non-gaussian acyclic model for causal discovery.
\newblock \emph{Journal of Machine Learning Research}, 7\penalty0 (10), 2006.

\bibitem[Spirtes(2013)]{spirtes2013calculation-t-separation}
Peter Spirtes.
\newblock Calculation of entailed rank constraints in partially non-linear and cyclic models.
\newblock In \emph{Proceedings of the Twenty-Ninth Conference on Uncertainty in Artificial Intelligence}, pages 606--615. AUAI Press, 2013.

\bibitem[Spirtes et~al.(2000)Spirtes, Glymour, and Scheines]{spirtes2000causation}
Peter Spirtes, Clark~N Glymour, and Richard Scheines.
\newblock \emph{Causation, prediction, and search}.
\newblock MIT press, 2000.

\bibitem[Spirtes et~al.(2013)Spirtes, Meek, and Richardson]{spirtes2013causal}
Peter~L Spirtes, Christopher Meek, and Thomas~S Richardson.
\newblock Causal inference in the presence of latent variables and selection bias.
\newblock \emph{arXiv preprint arXiv:1302.4983}, 2013.

\bibitem[Stram and Wei(1986)]{stram1986methodological}
Daniel~O Stram and William~WS Wei.
\newblock A methodological note on the disaggregation of time series totals.
\newblock \emph{Journal of Time Series Analysis}, 7\penalty0 (4):\penalty0 293--302, 1986.

\bibitem[Sullivant et~al.(2010)Sullivant, Talaska, and Draisma]{sullivant2010trek}
Seth Sullivant, Kelli Talaska, and Jan Draisma.
\newblock Trek separation for gaussian graphical models.
\newblock \emph{The Annals of Statistics}, 38\penalty0 (3):\penalty0 1665--1685, 2010.

\bibitem[Tiao(1972)]{tiao1972asymptotic}
George~C Tiao.
\newblock Asymptotic behaviour of temporal aggregates of time series.
\newblock \emph{Biometrika}, 59\penalty0 (3):\penalty0 525--531, 1972.

\bibitem[Vo(2009)]{vo2009regime}
Minh~T Vo.
\newblock Regime-switching stochastic volatility: Evidence from the crude oil market.
\newblock \emph{Energy Economics}, 31\penalty0 (5):\penalty0 779--788, 2009.

\bibitem[Weiss(1984)]{weiss1984systematic}
Andrew~A Weiss.
\newblock Systematic sampling and temporal aggregation in time series models.
\newblock \emph{Journal of Econometrics}, 26\penalty0 (3):\penalty0 271--281, 1984.

\bibitem[Xie et~al.(2020)Xie, Cai, Huang, Glymour, Hao, and Zhang]{xie2020generalized}
Feng Xie, Ruichu Cai, Biwei Huang, Clark Glymour, Zhifeng Hao, and Kun Zhang.
\newblock Generalized independent noise condition for estimating latent variable causal graphs.
\newblock \emph{Advances in Neural Information Processing Systems}, 33:\penalty0 14891--14902, 2020.

\bibitem[Xuan and Murphy(2007)]{xuan2007modeling}
Xiang Xuan and Kevin Murphy.
\newblock Modeling changing dependency structure in multivariate time series.
\newblock In \emph{Proceedings of the 24th international conference on Machine learning}, pages 1055--1062, 2007.

\bibitem[Zhang et~al.(2012)Zhang, Peters, Janzing, and Sch{\"o}lkopf]{zhang2012kernel}
Kun Zhang, Jonas Peters, Dominik Janzing, and Bernhard Sch{\"o}lkopf.
\newblock Kernel-based conditional independence test and application in causal discovery.
\newblock \emph{arXiv preprint arXiv:1202.3775}, 2012.

\bibitem[Zhang et~al.(2015)Zhang, Huang, Zhang, Sch{\"o}lkopf, and Glymour]{zhang2015discovery}
Kun Zhang, Biwei Huang, Jiji Zhang, Bernhard Sch{\"o}lkopf, and Clark Glymour.
\newblock Discovery and visualization of nonstationary causal models.
\newblock \emph{arXiv preprint arXiv:1509.08056}, 2015.

\bibitem[{\v{Z}}liobait{\.e} et~al.(2016){\v{Z}}liobait{\.e}, Pechenizkiy, and Gama]{vzliobaite2016overview}
Indr{\.e} {\v{Z}}liobait{\.e}, Mykola Pechenizkiy, and Joao Gama.
\newblock An overview of concept drift applications.
\newblock \emph{Big data analysis: new algorithms for a new society}, pages 91--114, 2016.

\end{thebibliography}
